\documentclass[aps,prx,reprint,twocolumn,superscriptaddress, amsmath,notitlepage,longbibliography,
amssymb]{revtex4-1}
\usepackage[T1]{fontenc}

\usepackage[utf8]{inputenc}
\usepackage{graphicx}
\usepackage[colorlinks=true,hypertexnames=false]{hyperref}
\usepackage{amsmath}
\usepackage{amssymb}
\usepackage{braket}
\usepackage{bbold}
\usepackage{times}
\usepackage{color}
\usepackage{xcolor}

\date{\today}

\newcommand{\pf}[0]{\operatorname{pf}}

\begin{document}

 \title{Classical simulation of parity-preserving quantum circuits}

\author{C.~Wille}
\affiliation{DAMTP, Centre for Mathematical Sciences, University of Cambridge, UK}
\author{S.~Strelchuk} 
\affiliation{Department of Computer Science, University of Oxford, UK}

\begin{abstract}
We present a classical simulation method for fermionic quantum systems which, without loss of generality, can be represented by parity-preserving circuits made of two-qubit gates in a brick-wall structure. We map such circuits to a fermionic tensor network and introduce a novel decomposition of non-Matchgate gates into a Gaussian fermionic tensor and a residual quartic term, inspired by interacting fermionic systems.
The quartic term is independent of the specific gate, which allows us to precompute intermediate results independently of the exact circuit structure and leads to significant speedups when compared to other methods. Our decomposition suggests a natural perturbative expansion which can be turned into an algorithm to compute measurement outcomes and observables to finite accuracy when truncating at some order of the expansion. For particle number conserving gates, our decomposition features a unique truncation cutoff reducing the computational effort for high precision calculations. 
Our algorithm significantly lowers resource requirements for simulating parity-preserving circuits while retaining high accuracy, making it suitable for simulations of interacting systems in quantum chemistry and material science.  Lastly, we discuss how our algorithm compares to other classical simulation methods for fermionic quantum systems.
\end{abstract}
\date{\today}
\maketitle

\section{Introduction}

Parity-preserving quantum circuits, encompassing the class of matchgate circuits are intensely studied in quantum information science due to their unique position at the intersection of quantum complexity and physical realizability~\cite{Valiant, knill2001fermionic,jozsa2008matchgates,brod2013computational,bravyi2002fermionic,bravyi2011classical}. Their efficient classical simulability under specific architectural constraints, such as nearest-neighbor connectivity, yields a powerful theoretical framework for studying the boundary between the classical and quantum theory of computation. This  makes them useful for benchmarking nascent quantum hardware~\cite{burkat2024lightweight} and validating the performance of early-stage quantum algorithms against their classical counterparts~\cite{boixo2018characterizing, oszmaniec2014classical}. Moreover, the direct mapping between these circuits and systems of non-interacting fermions establishes their relevance for simulating complex quantum phenomena arising in quantum chemistry and condensed matter physics, offering a potentially tractable route for exploring strongly correlated electron systems. 

Furthermore, the inherent parity conservation in these circuits offers significant advantages for quantum error detection. Errors that flip a single qubit directly violate the parity, making them readily identifiable. Efficient classical simulation techniques, such as those based on tensor networks, are essential for studying the behavior of these error-protected quantum systems and for designing effective decoding strategies. 

Classical algorithms for simulating quantum circuits are essential tools in understanding the emergence of complexity in quantum computations. Moreover, they are crucial for practical applications -- while reliable, large scale quantum computers are not yet widely available, advances in quantum many-body physics, quantum chemistry and material science heavily rely on numerical algorithms performed on classical computers. With both theoretical and practical questions in mind, in this work we devise a classical algorithm to simulate quantum circuits of parity-preserving two-qubit gates by mapping the circuit to a fermionic tensor network and study the performance of our algorithm as we vary the deviation of the fermionic tensor network from being Gaussian. 
 
Note, the restriction to parity-preserving gates does not restrict the computational power of the circuit, as any circuit can be cast into this form with a moderate encoding overhead (a factor two in the number of required qubits  \cite{Brod_2011}). 
 
 We study the circuits in terms of matchgates (MGs) and non-matchgates (non-MG). Similar to the setting of Clifford plus non-Clifford (e.g., T-gates), circuits composed of MG can be simulated efficiently on a classical computer and thus do not form a universal gate set. However, when augmented by any non-MG \cite{all_magic} -- typical examples being CZ, CPHASE (Cph) and SWAP gates -- the gate set becomes universal. In light of the diversity of quantum hardware implementations which vary substantially in their native gate sets, it is important to explore different partitions of universal gate sets into easy (classically simulatable, non-universal) and hard (universality enabling) gates. In addition to that, the framework of MGs plus non-MGs is natural for fermionic simulations.

Our algorithm is based on decomposing every non-MG into a sum of two components -- a (unitary) MG and a non-unitary projection and thus differs from the decomposition based on the Gaussian extent \cite{cudby2023gaussian} \footnote{The latter is defined for magic states \cite{all_magic} but can equally be applied to 2-qubit gates via Choi-Jamilkowski isomorphism.} used in Refs. \cite{ReardonSmith2024improvedsimulation,dias2023classical}. Instead, it is inspired by the structure of interacting fermion models. To be more precise, after mapping the circuit to a fermionic tensor network \cite{gu2010grassmanntensornetworkstates,PhysRevA.80.042333,PhysRevA.81.052338,Mortier_2025} via Jordan-Wigner (JW) transformation, each non-MG becomes a fermionic tensor that we decompose into a Gaussian fermionic tensor (corresponding to a MG) and a purely quartic term (corresponding to the projector). This decomposition provides the following advantages.

Firstly, it allows for precomputations that enhance computational efficiency. As the quartic term itself remains uniform across all non-MGs --  differing only by a prefactor specific to the gate -- we can compute numerically costly intermediate results independent of any particular non-MG. Those results can then be recycled to simulate the circuit for a whole range of non-MGs. This significantly reduces  the cost of simulating circuits for a whole parameter range -- a setting that is of particular relevance and routinely encountered in physics, for example, in the simulation of interacting systems, where one is interested in the dependency of physical quantities on the interaction strength.

Secondly, another feature of our approach is the ability to naturally perform approximate simulations. While exact computations generally require exponential runtime, as expected in quantum simulation, we focus on performing approximate computations to finite accuracy. This is achieved by casting the computation into the form of a series expansion which closely resembles the perturbation series commonly used in interacting fermion systems. Such a truncation allows us to reduce the runtimes while maintaining accuracy within a desired threshold.

Thirdly, while a similar perturbative expansion and truncation can be performed for other gate decompositions, we observe an additional advantage unique to our decomposition that comes into effect when dealing with particle number conserving gates. In these settings, there is a cutoff in the perturbative expansion -- depending on the initial state -- beyond which all terms vanish exactly and thus need not be computed. This feature is a direct consequence of the decomposition and leads to significant (albeit polynomial) reductions in the computational resources required. In particular, computations of high precision and for strong non-linearity (deviation from being a MG) profit from this effect.

The rest of this work is structured as follows. In Sec.~\ref{sec:parity} we discuss the mapping of parity-preserving circuits to fermionic tensor networks and introduce our gate decomposition. We establish the notion of simulating a quantum circuit used throughout this work and present our algorithm in Sec.~\ref{sec:algorithm}. In Sec.~\ref{sec:numerics} we study our algorithm numerically in two different settings -- time evolution in fermionic systems and random circuits. We conclude our work by a discussion and comparison to previous algorithms and speculate on future research directions in Sec.~\ref{sec:discussion}. 

\section{parity-preserving circuits}\label{sec:parity}
We consider quantum circuits of nearest-neighbor parity-preserving unitary (PPU) two-qubit gates. Note that with a simple encoding \cite{Brod_2011}, using two physical qubits to form one logical qubit, any circuit can be mapped to a PPU circuit. In addition, PPUs occur naturally in simulations of femionic systems. In the following we will discuss which instances of PPU gates are sufficient for universal quantum computation and, complimentary, which instances are efficiently simulatable on classical computers. For this, it is instructive to consider the explicit parametrization of any two-qubit PPU 
 $G(a,b)$ in terms of two matrices $a,b \in U(2)$ 
\begin{equation}G(a,b)=\begin{pmatrix}
	a_{11} & ~ &~ &a_{12} \\ & b_{11} & b_{12} & \\ & b_{21} & b_{22} & \\ a_{21}& & & a_{22} \label{eq:G}
\end{pmatrix}\;.
\end{equation}
If $\det A=\det B$, the gate is a MG and corresponds to a free or \emph{Gaussian} fermionic tensor \cite{Terhal_2002,Bravyi_2009}. Hence, $\gamma=\det A-\det B$, is a measure of  \emph{non-Gaussianity}. For both SWAP ($\gamma=2$) and CZ ($\gamma=-2$) $|\gamma|$ reaches its maximal value, while for CPHASE$(\phi)$ (Cph($\phi$)) and fSim$(\theta,\phi)$ gates $|\gamma|=2|\sin(\phi/2)|$ interpolates smoothly between $0$ and $2$.

\begin{figure}
\centering	
	\includegraphics[width=0.45\textwidth]{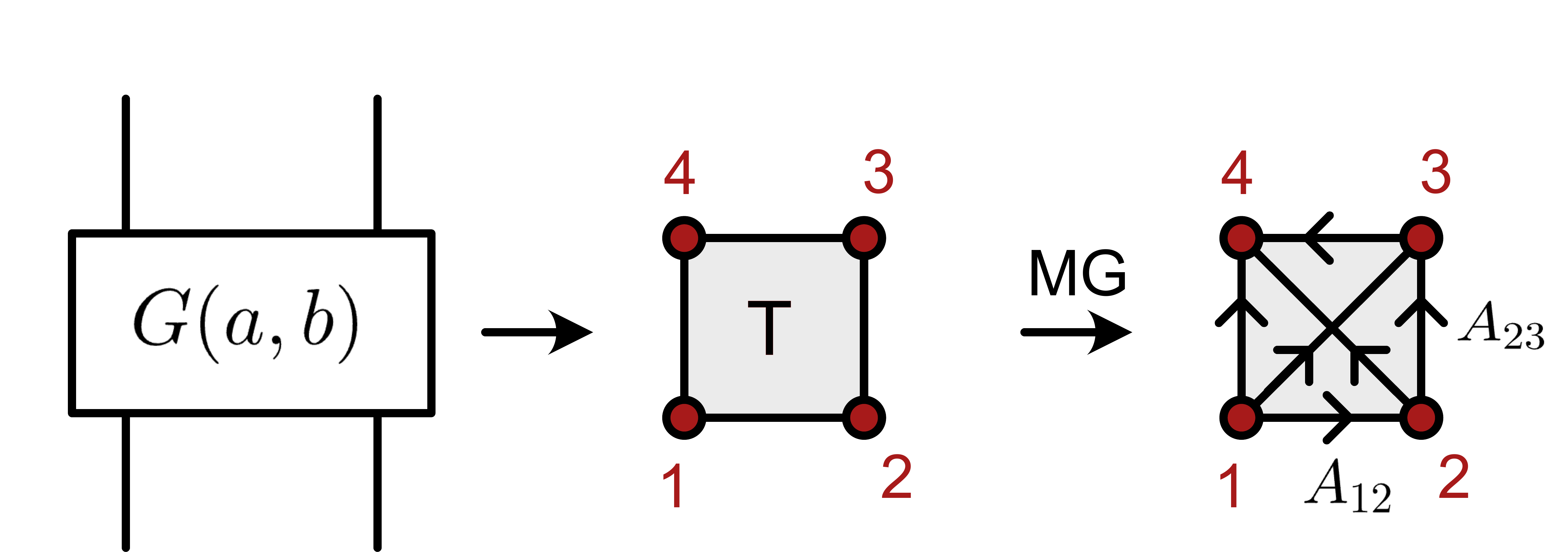}
	\caption{Mapping a PPU $G(a,b)$ to a fermionic tensor $T$ by associating fermionic modes to the indices and choosing an ordering. If the gate is a MG, its fermionic counterpart is a Gaussian tensor $T_\text{G}$ (cf. Eq.~\eqref{eq:Tdecomp}) and its generating matrix $A$ can be represented as a fully connected, directed, weighted graph.}
	\label{fig:map}	
\end{figure}

While circuits composed solely of MGs can be efficiently simulated classically \cite{Valiant,Terhal_2002,Brod_2016}, gates with non-vanishing non-Gaussianity enable universal quantum computation \cite{jozsa2008matchgates,Brod_2011} and prevent efficient classical simulation. The non-Gaussianity defined above is an alternative measure of the usefulness of a gate to perform quantum computation. In order to compare it to more established measures of non-Gaussianity, we recall that one can map any non-MG PPU to a 4-qubit magic state. Magic states are characterized by their (exact or approximate) Gaussian rank and their Gaussian extent. Concretely, the Gaussian extent $\xi$ is defined by expressing a state in an orthonormal basis of Gaussian states $\ket{\Psi}=\sum_i \alpha_i \ket{\Psi^{G}_i}$ and computing $\xi=|\sum_i \alpha_i|^2$ for the ideal (minimal $\xi$) decomposition. 

The extent $\xi$ and the non-Gaussianity $|\gamma|$ are closely related. In particular for fSim gates we have the following one-to-one correspondence $\xi(\phi)=1+|\gamma(\phi)|/2$. Both quantities come with unique decompositions of non-MG into sums of MG. While for the extent, the decomposition follows from minimizing $\xi$, a decomposition based on non-Gaussianity is more natural in the fermionic language, as shown below. 

Following the ideas in Ref.~\cite{Bravyi_2009}, we map general PPUs on a square lattice to fermionic tensors \cite{Lootens_2023,Wille_2024, Frank}. Under this map each PPU $G=\sum_{ijkl}\ket{i,j}G_{ij,kl}\bra{k,l}$ is mapped to a fermionic tensor $T=T_{ijlk} \theta_1^i \theta_2^j \theta_3^l \theta_4^k$, where $\theta_i$ are Grassmann variables~\cite{bravyi2004lagrangian}, representing the four fermionic indices (cf. Fig.~\ref{fig:map}) by identification of the matrix elements as $T_{ijlk}=G_{ij,kl}$. 

 It is then straightforward to show that after JW transform any PPU with $G_{00,00}\neq 0$ \footnote{If $G_{00,00}=0$ one needs to work with a slightly more complicated representation (cf. \cite{Bravyi_2009}), which is not discussed here.} can be decomposed into  a Gaussian tensor $T_\text{G}$ and a residual quartic term \cite{wille2024minimaltensornetworkfree}
\begin{equation}
T=T_\text{G}+ N \gamma \theta_1 \theta_2 \theta_3 \theta_4\;,\quad T_\text{G}=Ne^{\frac 1 2 \theta^T A \theta}\label{eq:Tdecomp} \;.
\end{equation}
Here, $A^T=-A$ is a $4\times 4$ matrix, $N \in \mathbb C$ is a 'normalization' and $\gamma \in \mathbb C$. The explicit expressions for $N,A$ and $\gamma$ are stated in App.~\ref{app:map}. If the tensor fullfils the MG condition, $\gamma=0$, it is Gaussian and its four-fermion entry $T_{1111}=\pf A/N$ is completely determined via its two-fermion entries $A_{ij}$.

The decomposition stated above is inspired by interpreting the quartic term weighted by $\gamma$ as an 'interaction' added to an otherwise free (Gaussian) fermionic system. It is non-orthogonal and presents an alternative to the minimal decomposition used to define the extent. To better understand the difference between the two decompositions, we consider a Cph gate and decompose it according to the extent $\xi$ or according to the non-Gaussianity $\gamma$
\begin{align} \label{eq:decomp}
	\gamma: \text{C-Ph}(\phi)=& e^{\frac 1 2 \theta^T A_\text{id} \theta} + a(\phi) \theta_1 \theta_2 \theta_3 \theta_4 \;,  \\[0.3cm]
	\xi: \text{C-Ph}(\phi)=& c_1(\phi) e^{\frac 1 2 \theta^T A_1(\phi) \theta} 
	+ c_2(\phi) e^{\frac 1 2 \theta^T A_2(\phi) \theta} \nonumber \;,
\end{align}
where $a(\phi)=1-e^{i\phi}$, $c_1 (\phi) = e^{\mathrm i \phi/4}  \cos (\phi/4)$, $c_2(\phi)=e^{\mathrm i \phi/4}  i \sin (\phi/4)$
and $A_\text{id}$, $A_{1}$ and $A_{2}$ are stated in  App.~\ref{app:As}.

Note, that for the $\gamma$-decomposition only the coefficient $a(\phi)$ is $\phi$-dependent. In contrast, for the $\xi$-decomposition, the generating matrices $A_1$ and $A_2$ are $\phi$-dependent as well. This is an important distinction between the two decomposition and implies the following fact. For the $\gamma$-decomposition we can perform the numerically costly part of simulating a circuit with one (or several) Cph gates independently of the angle $\phi$ and include the concrete value of $\phi$ at the very end of the computation with next to no numerical cost. This leads to a speed up for computations that involve a parameter sweep over $\phi$ based on precomputation. The details of this construction will become clear when we present our concrete algorithm to simulate a circuit in the following section.

\section{Algorithm}\label{sec:algorithm}
\begin{figure}\centering
	\includegraphics[width=0.48\textwidth]{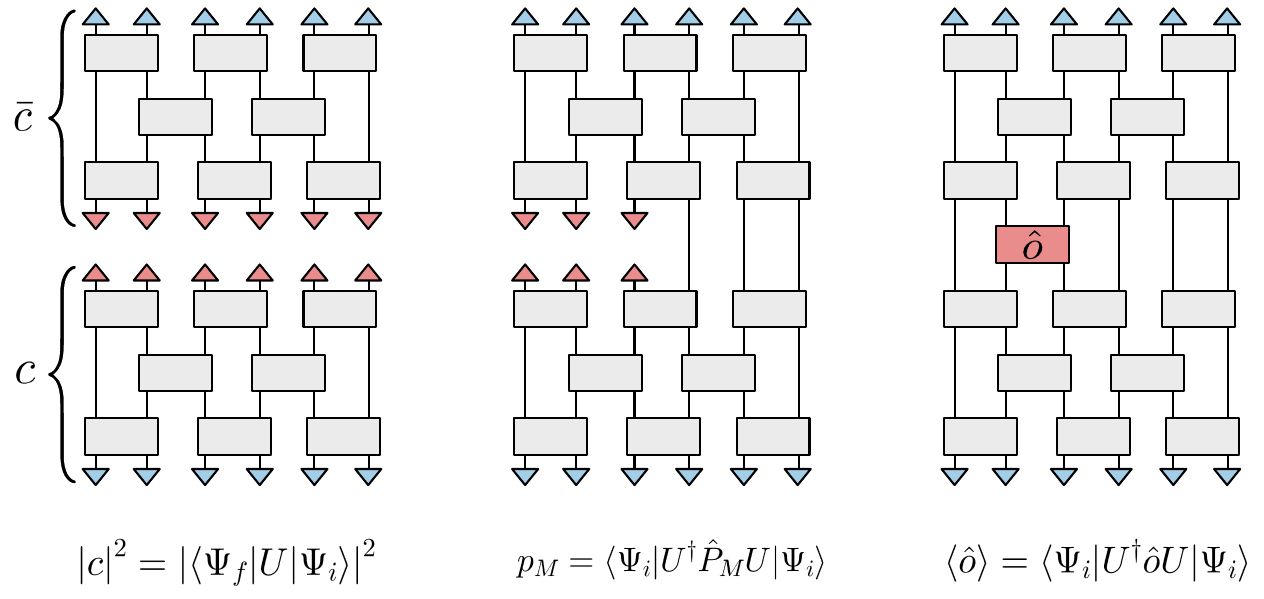}
\caption{Tensor networks yielding the probabilities of full qubit measurements (left), partial qubit measurements (center) and expectation values for local observables (right).}
\label{fig:c_and_related}
\end{figure}

There are various notions of what is meant by 'simulating a quantum circuit'. Our focus here is the (approximate) computation of the phase-sensitive overlap $c=\braket{\Psi_f|U|\Psi_i}$ of an initial state evolved under a PPU circuit $U$ with a final state. We focus on the case, where both states are parity even product states in the computational basis. The computation of $c$ corresponds to a tensor network (TN) contraction. This TN can be considered as an elementary building block for the computation of various other quantities (cf. Fig.~\ref{fig:c_and_related}). First of all, the probability of a projective all-qubit measurement $\hat P=\ket{\Psi_f}\bra{\Psi_f}$ is given by
 $|c|^2$ and can be inferred from $c$. To elevate this to a less-than-all qubit measurement one can either resort to sampling over a suitable basis for the non-measured qubits \cite{PhysRevLett.116.250501,dias2023classical}, or, alternatively, if one wants to avoid probabilistic methods, one can express the probability $p_M=\braket{\Psi_i |U^\dag P_M U | \Psi_i}$ for a  projective measurement $P_M$ as a tensor network, which features at maximum twice as many gates as the network $c$ \footnote{The network for $p_M$ can be slightly simplified by removing all unitaries outside the $P_M$-light cone.}. Similarly, one may consider expectation values $\langle o \rangle =\braket{\Psi_i|U^\dag  \hat o U | \Psi_i}$ of observables $\hat o$. These are again described by a TN containing at most twice as many gates as $c$. In the following, we will only study the elementary object $c$ and argue that for our method the runtime to approximately compute $p_M$ or $\langle o \rangle$ is not qualitatively different from the computation of $c$ except for number of gates. 

\subsection{Contraction via decomposition}

To compute $c$ we contract the corresponding TN. In order to take advantage of the efficient computability of planar MG TNs, we decompose each non-MG into two fermionic tensors as described above either using the $\gamma$- or the $\xi$-based decomposition. The contraction is then most easily expressed in the fermionic language, where we can represent it as a Grassmann integral. Concretely, the contraction of indices $i$ and $j$ corresponds to the integral $\int d \theta_{i} d \theta_{j}\, e^{\theta_{i} \theta_{j}}$ \cite{gu2010grassmanntensornetworkstates,Wille_2024} and the full contraction of a fermionic tensor network (fTN) on a square lattice is given by
\begin{equation}
c=\mathcal C^\text{f}(\{T\})=	\int (d \theta)_C \,e^{\frac 1 2  \theta^T C  \theta} \; T^1  \ldots T^N \;, \label{eq:contracted}
\end{equation}
where $T^1,\ldots T^N$ represent all fermionic tensors, $C=-C^T$ represents the signed adjacency matrix of a square lattice and $(d \theta)_C$ is a shorthand for the product of all
pairs of Grassmann variables to be integrated out. Note, that when mapping a bosonic to a fermionic TN, we need to make a consistent choice for the ordering of individual modes per tensor and the contraction order per index, which is captured by the directionality of the edges, stored in the sign structure of $C$. For our choice of ordering (cf. Fig.~\ref{fig:map}), the assignment of directions shown in Fig.~\ref{fig:lattice_arrows} in App.~\ref{app:JW} is a consistent choice \cite{Wille_2024}.

Let us recall, that the contraction of a planar MG TN can be performed efficiently via the computation of a Pfaffian. In particular, if all gates in the TN are MGs, the contraction above evaluates to
$$ c= \mathcal N \int (d \theta)_C \exp \left( \frac 1 2 \theta^T (A+C) \theta \right) = \mathcal N \pf (A+C) \;,$$
where $A=\oplus_i A^{(i)}$ and $\mathcal N =\prod_i N^{(i)}$ are the collections of $A$-matrices and normalizations $N$ determining the tensors $T$ via Eq.~\eqref{eq:Tdecomp}. 

\begin{figure}
	\centering
	\includegraphics[width=0.45\textwidth]{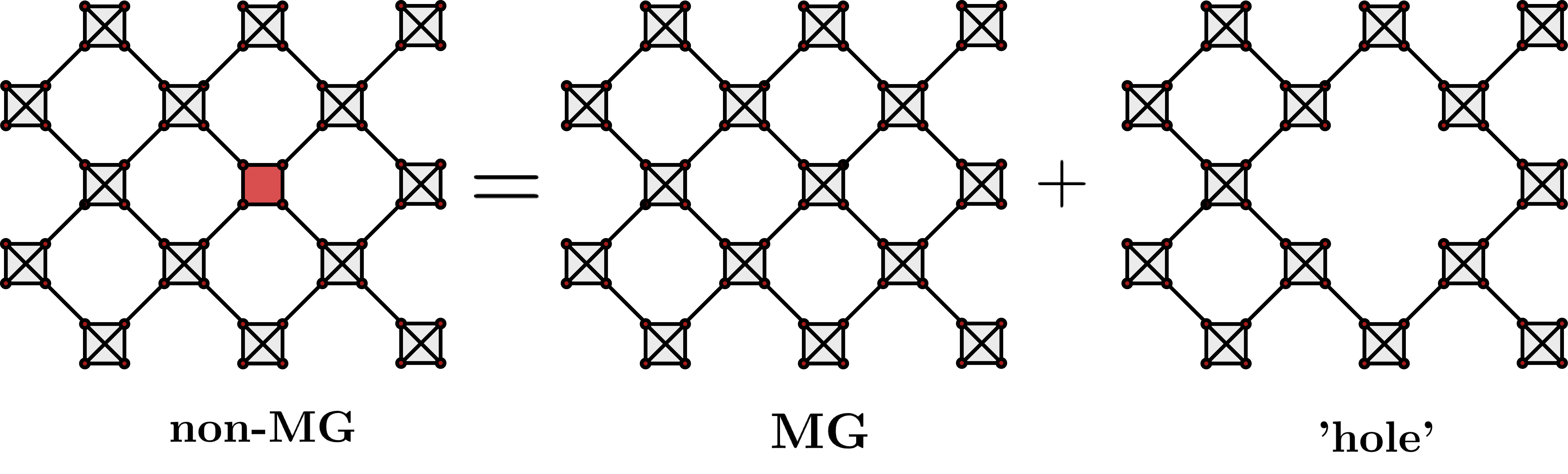}
	\caption{Decomposing a TN with a single non-MG into a sum of MG tensor networks -- one with the original topology plus one where the non-MG has been replaced by a hole.}
	\label{fig:deletion}
\end{figure}

We now decompose all non-MG PPUS into MGs. While the $\xi$-decomposition has been studied in Ref.~\cite{ReardonSmith2024improvedsimulation,dias2023classical}, we here focus on the $\gamma$-decomposition. Let us first consider a TN with a single non-MG specified by the triple $(N^{(i)},A^{(i)},\gamma^{(i)})$ at some position $i$ (cf. Fig.~\ref{fig:deletion}) and decompose this tensor according to Eq.~\eqref{eq:Tdecomp}. We obtain a sum of two tensor networks, one with the Gaussian tensor $T_\text{G}$ (MG) and one with a single four-fermion term inserted at position $i$, respectively. The four-fermion term projects the incoming four (fermionic) indices to the $\ket{1}$ state and thus decouples them completely. We can express this projection as the insertion of a hole into the TN (cf. Fig.\ref{fig:deletion}). 

After the decomposition both tensor networks are Gaussian and completely described by generating matrices. Denoting the first matrix by $M=\oplus_j A^{(j)}+C$, the second matrix, denoted $M|_i$ is obtained from $M$ by removing the rows and columns which correspond to the four fermionic modes at site $i$. The contractions of the TNs is then given by the weighted sum of the corresponding  Pfaffians and we obtain
\begin{equation}
c=\mathcal N \left(\pf M + \gamma^{(i)} \pf M|_i \right)\;,
\end{equation}
where $\mathcal N=\prod_j N^{(j)}$.

It is straight-forward to generzalize this expression to a setting with $m$ non-MGs. Here, we can express the contraction as a sum over all $2^m$ terms generated by the decomposition,
\begin{equation}c=\mathcal N \sum_{\{i_1,\ldots,i_m\}=0,1} \left(\prod_{j=1}^m \gamma_j^{i_j} \right) \pf M|_{\{i_1,\ldots,i_m\}}\;. \label{eq:c_decomp}
	\end{equation}
Here, $M|_{\{i_1,\ldots,i_m\}}$ is the matrix representing the TN with holes at all sites $j$ for which $i_j=1$.

\subsection{Runtime and Error}
Having in mind a setting where $\gamma_i\ll 1$, we sort the terms in the sum by the number of holes in the TNs $k=|\{i_1,\ldots,i_m\}|\equiv \sum_j i_j$. At order $k$, the terms are weighted by a product of $k$ $\gamma$-factors and there are ${m \choose k}$ Pfaffians to compute. For a circuit of size $s$ (total number of gates) the dimension of the matrices for which the Pfaffian is computed is $4(s-k)$ and the runtime per Pfaffian is $\mathcal O((s-k)^3)$. As expected \cite{Bravyi_2009,Brod_2011,dias2023classical,ReardonSmith2024improvedsimulation}, the exact computation requires $\sum_{k=0}^m {m \choose k}=2^m$ Pfaffian computations. However, for an approximate contraction $c_t$, we may truncate the sum at order $k_t<2^m$ and obtain the runtime 
\begin{equation}
	R(k_t,m,s)=\sum_{k=0}^{k_t} {m \choose k} \mathcal O((s-k)^3) \;. \label{eq:R}
\end{equation}
While the runtime is polynomial in the circuit size, the algorithm is costly in $k_t$ and $m$. More precisely, if the truncation order needed to achieve a certain accuracy scales linearly with $m$, the asymptotic runtime is exponential in $m$. In contrast, if the required truncation order is constant in $m$, the runtime is polynomial in $m$. 

For simplicity, we will now focus on the case $\gamma_i=\gamma$,
where the expansion is controlled by $\gamma^k$ and the truncation error $E=c-c_t$ is given by
\begin{equation}
	E(k_t,m,\gamma)= \mathcal N \sum_{|\{i_1,\ldots,i_m\}|=k>k_t}  \gamma^{k} \, \pf M|_{\{i_1,\ldots,i_m\}} \;.
\end{equation}

If $|\gamma|<1$, the coefficient in front of the Pfaffian sum decays exponentially in $k$. However, there is a combinatorial explosion of terms in the Pfaffian sum. Each Pfaffian is only bounded by $|\pf M|<1$ such that in the most adverse case their sum grows exponentially with $k$. However, we expect that in generic cases, the phases of the individual Pfaffians are largely independent and thus their sum at order $k$ corresponds to a random walk in two-dimension with variable step size smaller one. If this is the case the Pfaffian sums are expected to be orders of magnitude smaller than in the worst case. In particular, we expect that for $|\gamma|<1$, the exponentially decaying coefficients will dominate and allow us to achieve reasonable accuracies at low orders $k_t$.

\subsection{Faster evaluation through precomputation}
It is straight-forward to adapt the algorithm above and replace the $\gamma$-decomposition with the $\xi$-decomposition. In this case we essentially obtain a circuit-version of the building blocks used in Refs.~\cite{ReardonSmith2024improvedsimulation,dias2023classical} which work with MG circuits and magic states rather than non-MG gates. While the analog expression for $c$ in Eq.~\eqref{eq:c_decomp} for the $\xi$-based decomposition becomes somewhat more cumbersome, the structure remains identical. It is a sum of $2^m$ Pfaffians weighted by coefficients. In view of the arguments above, we expect that in a generic setting the coefficients of the $\xi$-based decomposition are smaller then the ones based on the $\gamma$-decomposition and thus, the number of terms needed to compute $c$ to a desired accuracy will be less. However, as noted earlier, all Pfaffians in the $\xi$-based decomposition depend explicitly on the specific non-MGs, while in the $\gamma$-based decomposition, the Pfaffians are independent of the specific gate and only the coefficients depend on it. 

This allows us to precompute the Pfaffian sums for any non-MGs and insert the specific form of the non-MGs afterwards. This substantially reduces the computational costs, whenever we need to compute $c$ for a range of non-MGs rather then for a single non-MG configuration. To illustrate this, consider the case of a circuit with $m$ Cph($\phi$) gates in a background of MGs. In this case, $\mathcal N=1$ and the $\gamma$-coefficient at order $k$ is given by $a(\phi)^k$ (cf. Eq.\ref{eq:decomp}). The matrices $M|_{i_1,\ldots,i_m}$ do not depend on $\phi$ and thus, we can precompute and store  their Pfaffian sums at order $k$ 
$$\text{PfSum}(k)=\sum_{|\{i_1,\ldots,i_m\}|=k} \pf M|_{i_1,\ldots,i_m}$$ 
in a list of length $k_t$. The final result $c$ is computed from this list in time linear in $k_t$ via $c=\sum_k a(\phi)^k \, \text{PfSum}(k)$.

This setting is of particular importance in the simulation of interacting fermionic systems, where  a circuit with uniform Cph$(\phi)$-gates occurs naturally in the Trotterized time evolution of the Hamiltonian and the angle $\phi$ is related to the interaction strengths. Thus, if one is interested in the dependence of quantities such as relaxation times or entanglement growth as a function of the interaction strength, a range of angles $\phi$ is naturally considered instead of just a single angle $\phi$. 

In the following section, we will investigate the performance of the algorithm above for this and other paradigmatic settings numerically and compare the $\gamma$- and $\xi$-based decompositions.

\section{Numerical Results} \label{sec:numerics}
We investigate the algorithm above numerically \cite{PfExpansion,Wimmer_2012} for two different kinds of circuits. First, we consider   circuits for the simulation of Trotterized Hamiltonian time-evolution and Floquet dynamics. These circuits feature an $AB$-pattern where layers of MGs and non-MGs alternate, and thus naturally have an extensive number of non-MGs. As a second benchmark for our algorithms we investigate random circuits composed of random MGs interspersed with non-MGs.

\subsection{Time evolution}
Having practical applications in mind, we consider how the algorithm performs for circuits that describe  discrete time-evolution. This setting includes Trotterized time-evolution of Hamiltonian systems, as well as Floquet dynamics, where a system is evolved periodically under two different Hamiltonians.

\begin{figure*}
	\includegraphics[height=0.2\textwidth]{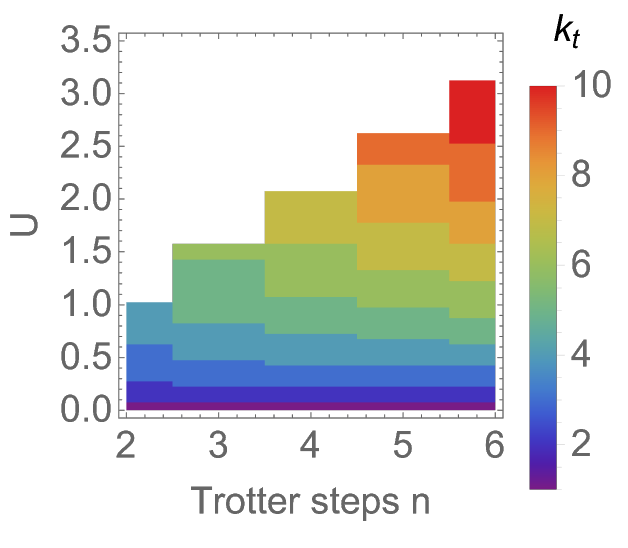} \hfill
				\includegraphics[height=0.2\textwidth]{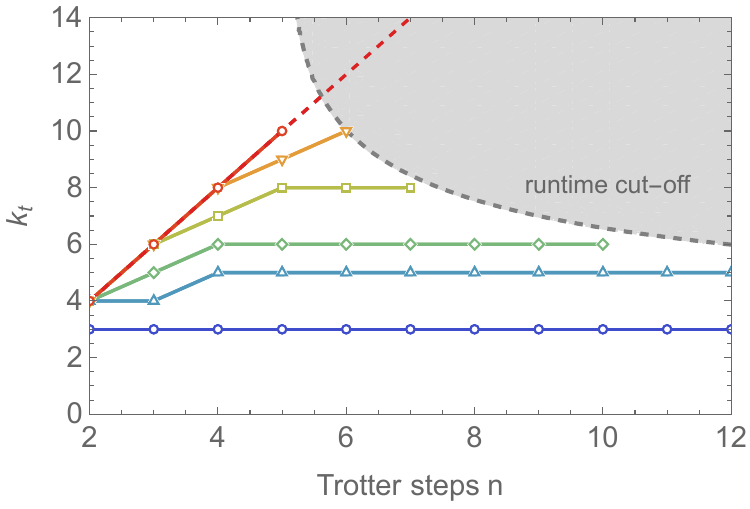}\hfill 
						\includegraphics[height=0.2\textwidth]{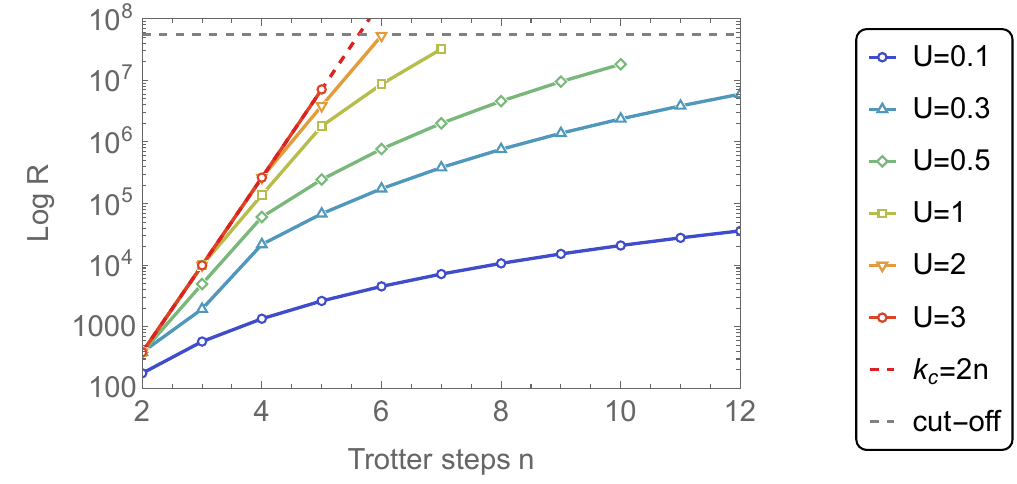}
	\caption{Trotterized time evolution of the tight-binding model in Eq.~\eqref{eq:H}. We perform the approximate computation of $c=\braket{\Psi_\text{h}|U_\text{T}(n)|\Psi_\text{h}}$ for $L=12$ qubits up to $0.01\%$ relative error and evaluate the truncation order $k_t$ (left, center) necessary to reach the given accuracy and runtime $R$ (right, log-scale) as a function of the interaction strength $U$ and Trotter steps $n$. For $k_c=2n$ (dashed lines) we obtain the exact result. All computations were terminated when the runtime exceeded a threshold.}
	\label{fig:Trotter}
\end{figure*}

We will focus on particle number conserving evolutions and in particular consider the following  tight-binding model as an example
\begin{equation}
	H=- t\sum_i  \left( \hat c_{i,1}^\dag \hat c_{i,2} + \hat c_{i,1}^\dag \hat c_{i+1,2}\right) + U \sum_{i} \hat n_{i,1} \hat n_{i,2} + \text{h.c.} \;. \label{eq:H}
\end{equation}
This Hamiltonian describes a 1D chain of spinless fermions with a two-site unit cell. Here, $t$ is the hopping amplitude and $U$ is the strength of the density-density interaction within the unit cell.

This model is similar to the Hubbard model for spin-full fermions with the difference that a two-site unit cell replaces the spin-degree of freedom. W.l.o.g. we set $t=1$ and consider its Trotterized time-evolution $U_\text{T}(U,n)$ for a total time $T=\Delta T n$ divided into $n$ Trotter steps. For each time step $\Delta T$ the circuit consists of three layers -- the first two layers account for intra and inter unit cell hopping and are Gaussian fSim-gates with  $\theta=t \Delta T$ and $\phi=0$. The third layer captures the density-density interaction and is given by CPh gates with $\phi=U \Delta T$. We can combine layer two and three into one layer of gates fSim$(t \Delta T,U \Delta T)$ and thus arrive at an $AB$-pattern where a MG layer of fSim$(t \Delta T,0)$ is followed by a non-MG layer of fSim$(t \Delta T,U \Delta T)$.

\paragraph{Precomputation.}
As mentioned in the last section, for our algorithm based on the $\gamma$-decomposition the Pfaffian sums can be computed independently of the angle $\phi$ for any Cph gate. This extends to the angle $\phi$ in fSim gates as well. Consequentially, we can precompute the list of Pfaffian sums for any computation $c(U,n)=\braket{\Psi_i | U_T(U,n)|\Psi_f}$ independent of the parameter $U$ and obtain the final result in linear time from the precomputed list. In practice, this amounts to adding and multiplying $\mathcal O (s)$ numbers and has negligible computation time. Thus, the precomputation essentially collapses the calculation time of an entire $U$-interval to that of a single $U$-point. This speed up is most significant in cases where a $U$-interval needs to be resolved with high precision.
Note, that in the $\phi$-decomposition such a precomputation is not possible, as all Pfaffians depend on $U$.
 
However, we should point out that at least in the simulation of Trotterized time evolution, the precomputation advantage is limited by the fact that one should not increase $U$ arbitrarily while keeping the number of Trotter steps $n$ unchanged. This is because the circuit $U_T$ approximates the true time evolution up to a Trotter error which is determined by $U/n$. Thus, to ensure a reasonable Trotter error $U/n$ must not become too large. However, with the precomputation method in mind, one can choose $n_\text{max}$ according to the maximal value $U_{\text{max}}$ that one is interested in and obtain all results for $U<U_\text{max}$ from the computation $c(U_\text{max},n_\text{max})$ in one shot.

A circuit with the same structure as the above also arises in Floquet settings. In this case, the angle $\phi$ in the fSim gates need not be small to keep a Trotter error at bay, but can be unbounded. In this case, there is no restriction to the precomputation speed up of collapsing a full $\phi \in [0,2\pi)$ interval to a single point.

\paragraph{Scaling.}
Apart from the precomputation scheme detailed above, let us investigate the runtime scaling for a single computation with fixed $U$. As already mentioned $t \Delta T$ and $U \Delta T$ are required to be small in order to avoid large Trotter errors which is ensured by increasing the number of Trotter steps $n$. The number of non-MGs is proportional to the number of Trotter steps $n$ and thus, the simulation cost of exact contractions increases exponentially with $n$. For approximate contractions, the runtime scales with the truncation order. 

Note, that in our setting however, the effect of increasing $n$ is two-fold --  while increasing the number of Trotter steps increases the number of non-MG, it simultaneously decreases the angle $\phi=U/n$ of the CPh gates. While the former increases the computational cost, the latter is beneficial for approximate calculations as we expect that a smaller $\phi$-value allows us to approximate the circuit with a lower truncation order $k_t$. As a consequence we expect the truncation order needed to achieve a truncation order that is of the same order as the Trotter error to increase only moderately with $n$. We test this intuition numerically.

Fig.~\ref{fig:Trotter} shows the truncation order and runtime for the $\gamma$-based decomposition to approximate $c=\braket{\Psi_f|U_\text{T}|\Psi_i}$ to a desired accuracy. As the gates are particle number conserving, we here consider initial and final states with the same number of particles. In particular we choose $\Psi_{i,f}$ to be states at half-filling. We fix the time to be $T=1$ and investigate the scaling of $k_t$ as a function of the interaction strength $U$, the number of Trotter steps $n$ and the boundary conditions (the initial and final states). To limit the Trotter errors, we restrict the analysis to $(U,n)$ for which $\phi<\pi/6$ (arbitrary choice). Generically, we observe that there exists a regime of comparably small U and low accuracy, for which 
the truncation order stays constant in the number of Trotter steps and accordingly, the runtime scales polynomially with $n$. For larger values of $U$ and higher accuracies, the truncation order grows linearly, leading to an exponential increase of the runtime as expected.

\paragraph{Exact cut-off.}
We observe, that while the total number of non-MG $m$ is given by $m=n(L/2-1)$, where $L$ is the number of qubits, the Pfaffians in Eq.~\eqref{eq:c_decomp} vanish for all $k>k_c$ larger than a cut-off $k_c<m$ that depends on the initial and final states. The existence and magnitude of $k_c$ follows from the following observation. At each order $k$ we introduce $k$ projections $\ket{1,1}\bra{1,1}$ into the TN. This corresponds to enforcing a certain particle number at intermediate time-steps in various regions of the TN which we can think of as a spatio-temporal particle number pattern. As the individual gates conserve the particle number, the accessible patterns are heavily constrained and depend on the initial state. For the half-filling states $\ket{\Psi_\text{h,e}}$, we find $k_c(\Psi_\text{h})=(L/4-1)n$ and $k_c(\Psi_\text{e})=L/4(n-1)$, which is about half of the total number of MGs.

\begin{figure}
	\centering
	\includegraphics[width=0.35\textwidth]{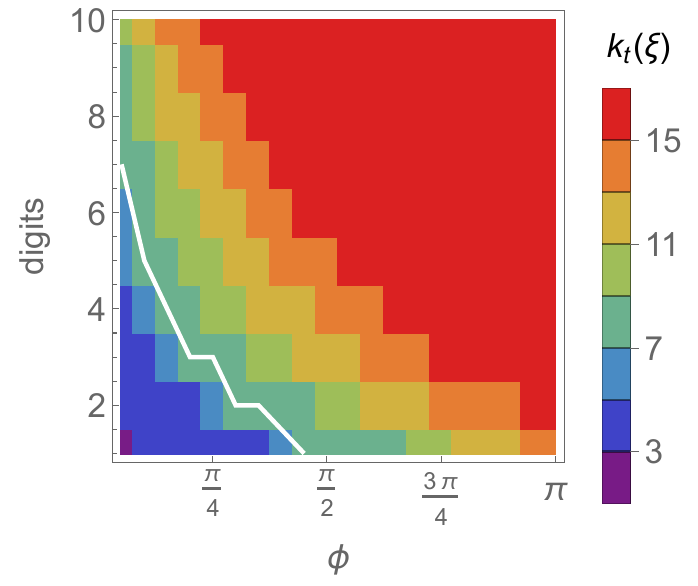}
	\caption{Floquet setting. Truncation order $k_t$ for the $\xi$-decomposition needed to approximately compute 
	$c=\braket{\Psi_\text{h}|U(\theta,\phi)|\Psi_\text{h}}$ as a function of $\phi$ and the desired accuracy measured in the numbers of digits for $l=6, d=3, \theta=0.1$. The white line marks $k_t(\xi)=k_c$. To the right of it the $\gamma$-decomposition outperforms the $\xi$-decomposition.}
	\label{fig:Floquet}
\end{figure}

Note, that this exact cut-off only occurs in the $\gamma$-based decomposition as it is tied to the projection onto fixed particle number states. In the $\gamma$-based decomposition, the exact calculation needs to be performed with $k_c=m$. Thus, for all settings, where a high truncation order is needed -- high accuracy calculations for large interaction strengths -- the $\gamma$-based decomposition outperforms the $\xi$-based decomposition. Fig.~\ref{fig:Floquet} shows such a scenario. To explore a wider range of non-Gaussianity in the individual gates, we relax the setting from Trotterized time evolution to general unitary evolution $U(\theta,\phi)$ that consist of a layer of Gaussian gates fSim($
\theta,0$) followed by a layer of non-Gaussian gates fSim($\theta,\phi$). This more general circuit would for example be realized in particle number conserving Floquet dymamics with a non-interacting and an interacting Hamiltonian. It reduces to the Trotterized Hamiltonian dynamics of the tight-binding model above for $\theta,\phi \ll 1$. We now compare the result of approximately computing $c=\braket{\Psi_\text{h}|U(\theta,\phi)|\Psi_\text{h}}$ for the two decompositions. To do so, we again calculate the truncation $k_t$ order needed to achieve a certain accuracy and explore how it scales with $\phi$ and the accuracy (measured in number of digits) for the two decompositions, respectively. 

Generically, we find that for settings in which the truncation order of the $\gamma$-based decomposition is below the exact cut-off, $k_t(\gamma)<k_c$, the $\xi$-based decomposition performs better, i.e, we have $k_t(\xi)<k_t(\gamma)$. However, for large $\phi$ and high accuracy calculations, $k_t(\xi)$ approaches and exceed $k_c$. For all these settings, the $\gamma$-decomposition outperforms the $\xi$-decomposition. Fig.~\ref{fig:Floquet} exemplifies this mechanism and  shows $k_t(\xi)$ needed to compute $c=\braket{\Psi_\text{h}|U(\theta,\phi)|\Psi_\text{h}}$ as a function of $\phi$ and the desired accuracy measured by the number of digits. The contour $k_t(\xi)=k_c$ (white line) divides the $(\phi,\text{digits})$-plane into the regions $k_t(\xi)<k_t(\gamma)$ and $k_t(\xi)>k_t(\gamma)=k_c$. In the latter region, the runtime for the $\gamma$-decomposition is constant throughout the region and shorter then the runtime of the $\xi$-decomposition, which still increases with $\phi$ and the number of digits until it saturates at the maximal values $k_t(\xi)=m$.

\subsection{Random Circuits}

\begin{figure}[t]
\centering
	\includegraphics[height=0.23\textwidth]{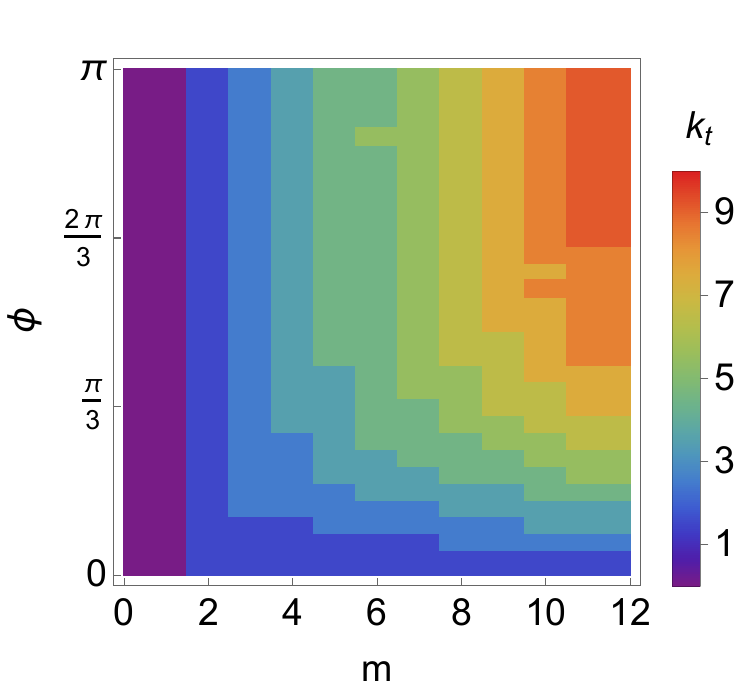}%
		\includegraphics[width=0.23\textwidth]{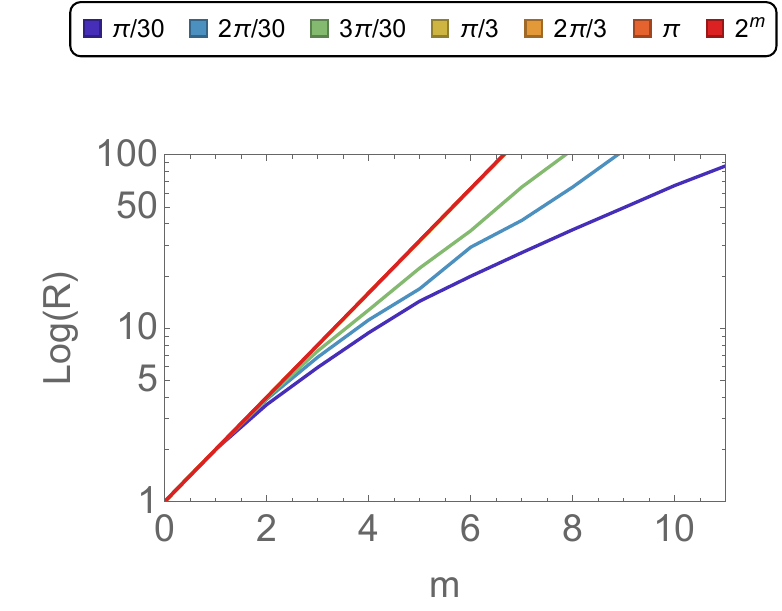}
	   \\
	\includegraphics[height=0.22\textwidth]{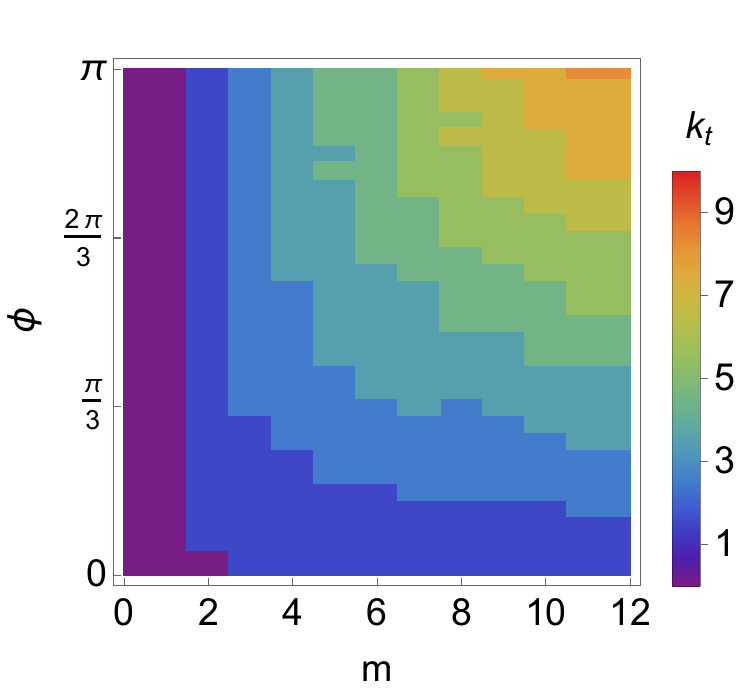}
	\includegraphics[width=0.23\textwidth]{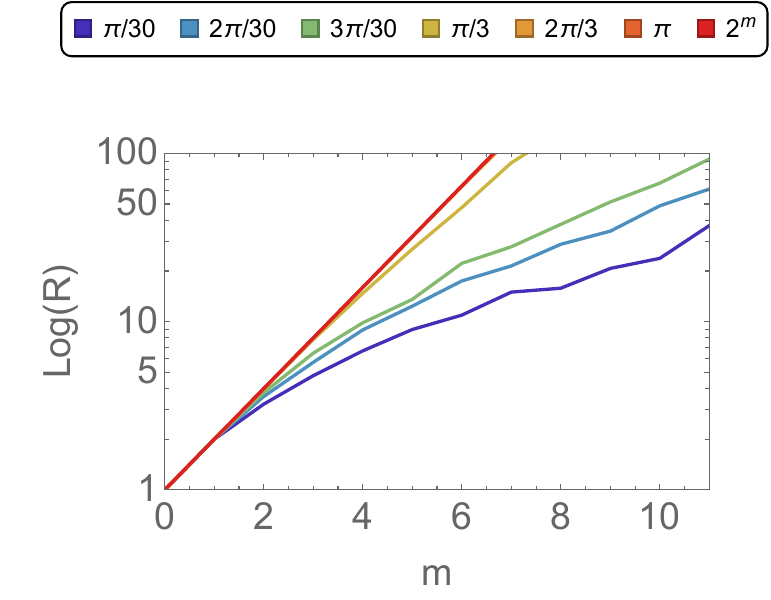} 
	\caption{Random circuits -- $\gamma$ (top) vs $\xi$ (bottom) based decomposition. Left: Truncation order $k_t$ needed to compute $c$ within relative error of $1\%$ for a circuit containing $m$ C-Ph gates with angle $\phi$, averaged over 100 samples. Right: Corresponding runtimes in logarithmic scale. Circuit depth $d=20$, number of qubits $l=20$.}
	\label{fig:random}
\end{figure}

\begin{figure}
	\centering
\includegraphics[width=0.35\textwidth]{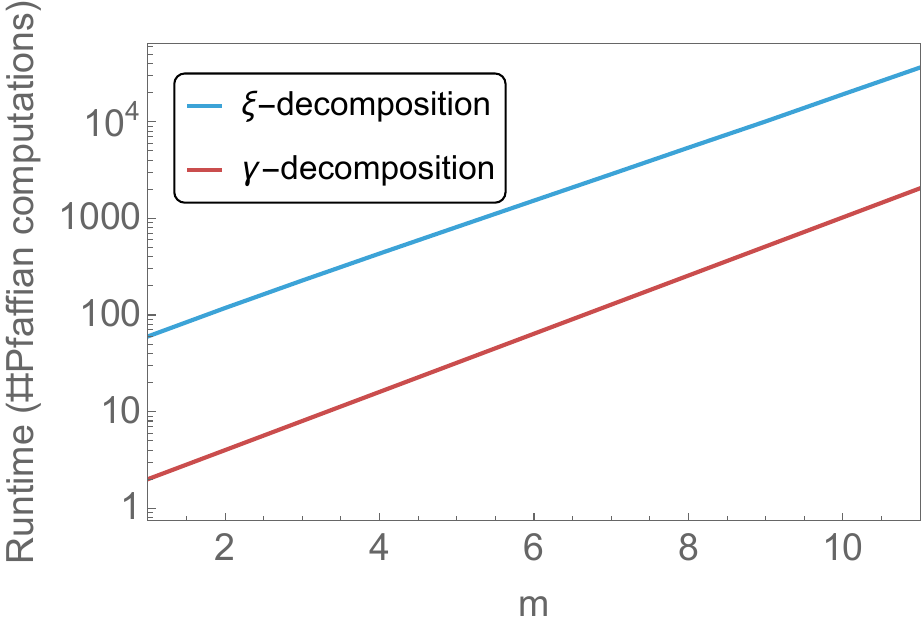}
\caption{Runtime to compute an entire $\phi$-column in Fig.~\ref{fig:random} for the $\gamma$-decomposition and the $\xi$-decomposition (used in \cite{dias2023classical}), respectively.}
\label{fig:compare}
\end{figure}

In addition to the time evolution circuits, we also test our algorithms on random circuits. Concretely, we investigate a circuit $U$ of random MGs, where we choose $m$ gates at random and replace them (i) by CPh($\phi$)-gates and (ii) by Haar random PPUs. As a benchmark for our algorithm we again compute $c=\braket{\Psi|U|\Psi}$ for $\ket{\Psi}=\ket{0}^{\otimes L}$ to finite accuracy and investigate how the truncation order $k_t$ necessary to achieve the given accuracy scales with the number of non-MGs $m$ and the (average) value of $|\gamma|$. The truncation order is in one-to-one correspondence with the runtime (cf. Eq.~\eqref{eq:R}). For large circuit sizes and a comparably small number of non-MG, we have $s\ll k_t$ and can approximate $R(k_t) \simeq C_1 \sum_{k=0}^{k_t} {m \choose k}$, where the constant $C_1$ depends on the circuit size, but not on $m$.

\paragraph{Random MG circuit with additional Cph gates.}
We compare the algorithms based on the $\gamma$- vs. the $\xi$-decomposition. Fig.~\ref{fig:random} presents the results. We observe that the scaling of $k_t$ as a function of $m$ and $\phi$ is  qualitatively similar for both decompositions. For large values of $\phi$, corresponding to $|\gamma|>1$, we observe a linear increase of the truncation order $k_t$ with $m$ which results in an exponential runtime. In contrast, in the small $\phi$-regime, where $|\gamma| \ll 1$, the growth of $k_t$ with $m$ is significantly reduced and constant over wide regions of $m$ leading to polynomial runtimes in $m$.

Comparing the $\gamma$- and the $\xi$-based decomposition, we find that the $\xi$-based decomposition outperforms the $\gamma$-decomposition for any given point in the $(m,\phi)$-plane. However, in the $\gamma$-based decomposition, we can again resort to precomputation and compute $c$ independent of $\phi$. Any concrete computation can then be calculated from the precomputed list in time linear in $m$. To illustrate the speedup gained by this method we compare the runtimes for computing entire $\phi$-columns. While the algorithm using the $\gamma$-decomposition can resolve $\phi$ with largely unlimited precision, we need to work with finite resolution of the $\phi$-interval in the $\xi$-decomposition as every point needs to be computed individually. The speedup gained from precompution using the $\gamma$-decomposition is then largely determined by the resolution as the individual runtime differences become negligible. To be precise, for the resolution of 30 points per interval with which Fig.~\ref{fig:random} was generated, the $\gamma$-decomposition is about 30 times faster as illustrated in Fig.~\ref{fig:compare}.

In addition, we note that while the runtime of both decompositions mainly depends on the truncation order, for the $\gamma$-based decomposition the size of the effective circuit is reduced at each order. While for $m\ll s$, this effect is marginal and has been neglected in the computation of the runtimes, in a setting where $m \sim s$, this effect becomes relevant and might lead to an advantage of the $\gamma$ over the $\xi$-based decomposition.

\begin{figure}
\centering
				\includegraphics[width=0.4\textwidth]{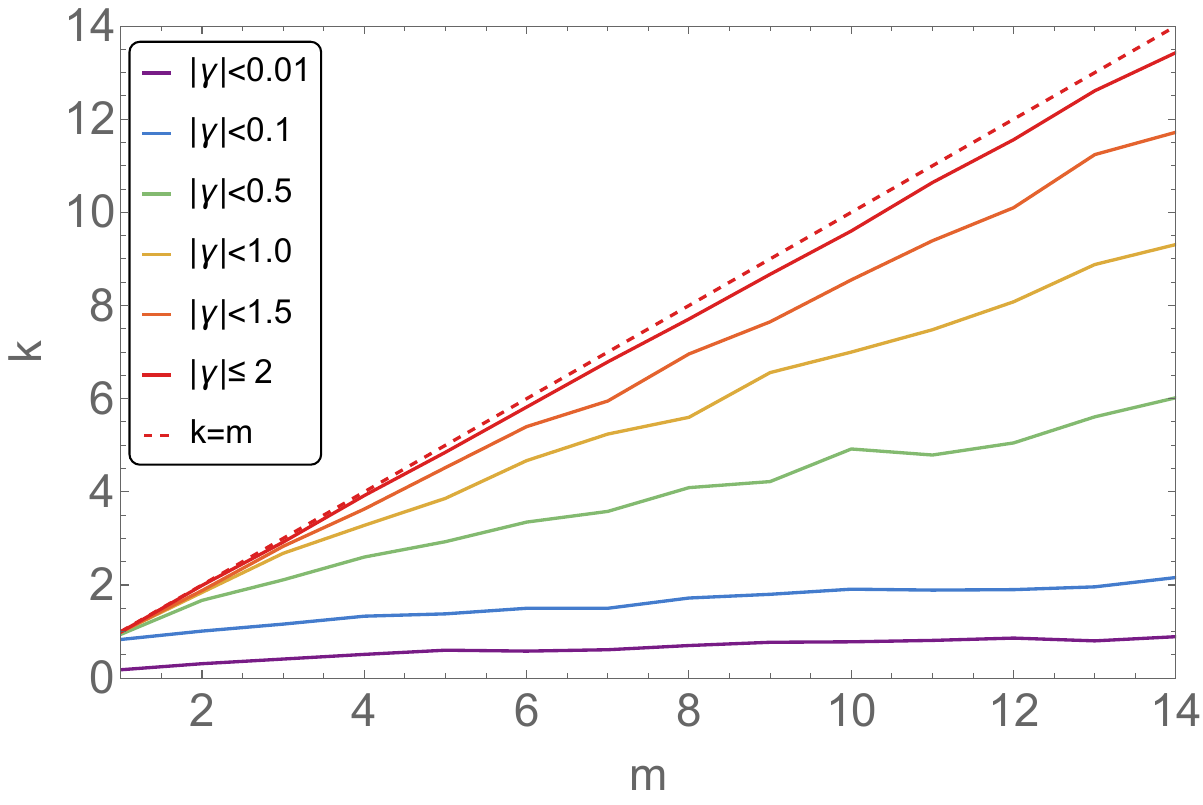} 			\includegraphics[width=0.42\textwidth]{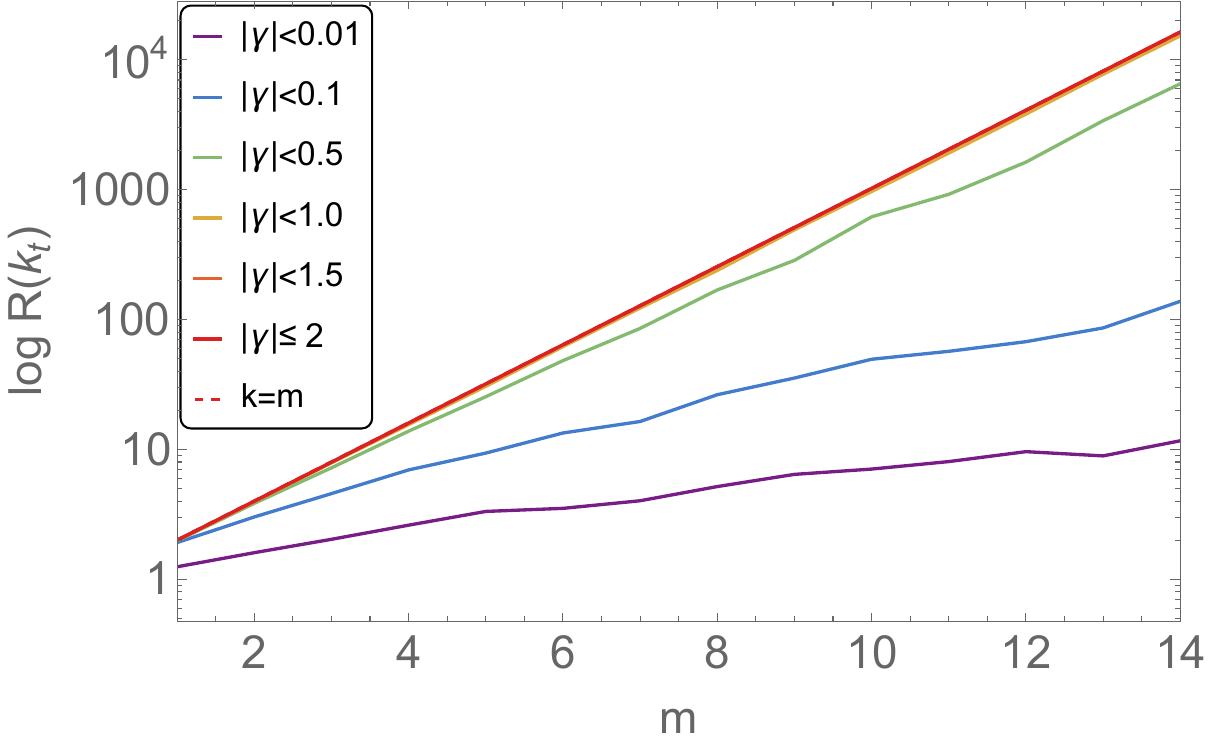}
	\caption{Truncation order and runtime for $m$ random PPUs drawn from a distribution with a $|\gamma|$-cut off. System size $d=l=6$, 100 samples, accuracy goal $1\%$ relative error.}
	\label{fig:r1}
\end{figure}

\paragraph{Random MG plus random PPUs.}
Replacing the Cph gates by random PPUs we observe a similar situation. Here, we generate Haar random PPUs which come with a particular $\gamma$-distribution. To compare different $\gamma$-distributions, we introduce a cut-off $\gamma_c$ and disregard all PPUs with $|\gamma|>\gamma_c$. The average $k_t$ needed to achieve a given accuracy in the presence of $m$ random PPUs chosen from a distribution with cut-off $\gamma_c$ depends on the average $\langle |\gamma| \rangle$ in much the same way as $k_t$ depends on $|\gamma|$ for a circuit with additional Cph gates as shown in Fig.~\ref{fig:r1}. In particular, we observe a transition from exponential to polynomial runtime in $m$ when $\gamma_c$ approaches sufficiently small values.

\section{Discussion and Outlook}\label{sec:discussion}
In this work we presented a classical algorithm for simulating quantum circuits of parity-preserving two-qubit gates in a nearest-neighbor (brick wall) architecture. More specifically our algorithm computes the probability to transition between two given computational basis states via a unitary circuit -- or equivalently, the probability of a full qubit measurement outcome (in the computational basis).

  This computation is expressed as a TN contraction which -- after mapping to a fermionic TN -- we evaluate in terms of a Grassmann integral. To calculate this integral we employ a perturbation expansion that is inspired by weakly interacting fermionic systems. Concretely, we decompose each tensor into a Gaussian tensor (equivalent to a MG) and a purely quartic term (equivalent to a $\ket{11}$-projection) and obtain a sum of TNs. The quartic term amounts to the deletion of the tensor from the network and thus we obtain a sum of Gaussian fermionic TNs with a varying number of holes. The individual contractions evaluate to Pfaffians which can be computed in polynomial time (in the number of gates). 
  
  \subsection{Summary of main features}
  For exact computations, each parity-preserving gate which is not a MG is decomposed according to the above scheme, and hence we need to compute an exponential number of Pfaffians. However, for approximate calculations we can employ a truncation scheme where we drop terms with more than a certain number of holes. This corresponds to limiting the amount of interaction terms that are taken into account simultaneously. Numerical simulations show that this truncation scheme works well in the regime where non-MG are sparse or deviate only weakly from being a MG, i.e., for circuits whose overall deviation from being Gaussian is weak.
  
  Besides the truncation scheme, which can be used to trade a moderate accuracy loss for a significant computational speedup, our algorithm features a precomputation advantage based on the fact that the second component in the decomposition does not depend on the specific gate (only its coefficient does). Concretely, for a single non-MG in a MG background, the Pfaffian corresponding to the second term in the decomposition can be computed once, stored and reused for all computations with an arbitrary non-MG at the respective site. This feature is particularly useful in uniform settings, where all non-MG in the circuit have identical Gaussian components as, for example, for CPh gates. In this case, not only the individual Pfaffians for the second component terms, but all Pfaffians can be recycled. If all CPh gates have the same angle $\phi$, we can go one step further and compute the entire Pfaffian sum for a $k$-hole term. Having performed one full computation for a given angle $\phi$, all following computations for different angles $\phi'$ can then be computed in linear time (in the number of CPh gates). 

In the case where the circuit does not only preserve parity but also the particle number, we observe that the series expansion terminates earlier than na\"ively expected. This exact cut-off depends on the particle number distribution of the initial state and the chosen measurement outcome. In the examples considered, this effect reduced the computational cost by a factor two, but it is straightforward to generate examples, where the Pfaffians corresponding to $\ket{11}$-projections at a large number of sites trivially vanish, reducing the number of Pfaffians to be computed by more then a factor of two. To see this, we consider computing the probability of two neighboring particles on the right edge of a chain to remain in place. The particles can only travel one site per time-step. As a consequence, all $\ket{11}$-components of gates outside the particles' light-cone vanish, and by tuning the circuit depth vs width we can reduce the fraction of non-MG for which both the Gaussian and the $\ket{11}$-component need to be considered arbitrarily.

Note that this mechanism relies on the gate-independence of the second component in the non-MG decomposition. In particular, decompositions into a sum of two MGs based on the Gaussian extent (referred to as $\xi$-decomposition throughout this text) do not feature this effect and show no reduction of the computational cost for particle number conserving circuits. The same applies to the precomputation advantage stated previously. It is these two features which -- depending on the concrete task at hand -- render our algorithm favourable compared to a decomposition based on the Gaussian extent. 

\subsection{Relation to recent work}

Following up on this observation, we now discuss in more depth how our algorithm compares to other recent works \cite{Hakkaku_2022,mocherla2023extending,ReardonSmith2024improvedsimulation,dias2023classical,cudby2023gaussian,projansky2024,miller2025simulationfermioniccircuitsusing}.

Among the aforementioned, our algorithm is most closely related to the works by \cite{ReardonSmith2024improvedsimulation} Reardon-Smith et al. and Dias and Koenig \cite{dias2023classical} in that it is formulated in fermionic language and based on the decomposition of non-Gaussian gates (or equivalently non-Gaussian magic state inputs) into a sum of two components. However, both works use a decomposition based on the Gaussian extent ($\xi$-decomposition), whereas we employ a different decomposition. While Reardon-Smith et. al. \cite{ReardonSmith2024improvedsimulation} do not employ any pruning techniques, and thus produce an algorithm that scales directly with the extent (which itself scales exponential in the number of non-Gaussian gates), Dias and Koenig \cite{dias2023classical} consider truncations based on the extent which can outperform our algorithm for a particular single computation, but does not feature the precomputation advantage of the exact cut-off that occurs in the particle number conserving settings. 

The algorithm by Mocherla et al. \cite{mocherla2023extending} is distinct from our work and that of Refs.~\cite{dias2023classical,ReardonSmith2024improvedsimulation} in that the authors focus on the simulation of expectation values of single Pauli measurements and consider product state inputs. Their algorithm uses the JW transform to map back and forth between Pauli strings and Majorana operators and tracks the evolution of Majorana polynomials under non-Gaussian gates. In each update a non-Gaussian operation maps a single monomial to a sum of monomials. The number of monomials with non-vanishing coefficient is referred to as the Pauli rank. The authors discuss a truncation scheme based on removing the terms with the smallest coefficients prior to the next update. A very similar algorithm has been proposed recently by Miller et al. \cite{miller2025simulationfermioniccircuitsusing} with the difference of a more elaborate truncation scheme.

In all previously mentioned works, and in contrast to ours, the truncation schemes operate on a sequential level, meaning prior to each update a truncation may or may not be employed. A potential pitfall of these schemes is that -- thinking in a path integral analogy -- trajectories which might be highly relevant in a later stage of the evolution (dominating the measurement outcomes) may be  removed early on, and thus lost entirely. This effect can be observed in sequential tensor network contractions, where the order of contractions and truncations is a crucial factor for the overall accuracy. In our algorithm, all non-Gaussian gates are treated on equal footing, irrespective of their temporal position in the evolution. It is conceivable that this is advantageous in certain settings, and it is left for future research to investigate this in more depth.

\subsection{Future directions}
In this work we have focused on full qubit measurements, which are easier to perform than partial qubit measurements. To see this, consider Fig.~\ref{fig:c_and_related} and observe that a full qubit measurement following a size-$s$ circuit corresponds to the contraction of a size-$s$ tensor network. In contrast, the network corresponding to a partial measurement is of size $2s$ (see Fig.~\ref{fig:c_and_related}). With runtimes exponential in the number of gates, this circuit is thus significantly harder to simulate. A na\"ive way to adapt our algorithm to this setting would be to insert a full set of basis states for the unmeasured qubits and treat every component as a full qubit measurement. However, this procedure scales as $2^l$, where $l$ is the number of non-measured qubits. Alternatively, one may consider sampling over the unmeasured qubits using intermediate states from an 'easy' set. In our case, this would correspond to a basis of Gaussian states. These sampling methods have been extensively studied in Ref.~\cite{ReardonSmith2024improvedsimulation} and can be combined with our approach in a straightforward fashion.

Instead of summing over intermediate states, one may also apply our algorithm directly to the size $2s$-circuit. This can be made more efficient by exploiting causal cone arguments to shrink the size of the non-trivial part of the circuit. The detailed comparison of the runtime for approximate computations via this direct approach vs sampling based method is left for future work. 

Our algorithm is based on two main ideas -- the computation of a tensor network contraction via a Grassmann integral and the decomposition of a parity-preserving gate in terms of a Gaussian fermionic tensor and a purely quartic term. While we have studied both of these ingredients together, we assume that it will be fruitful to investigate them separately and in conjunction with other techniques. More precisely, the decomposition scheme can be combined with other contraction and evolution methods. For example, it could just as well be employed in the computation of expectation values and the evolution of mixed states, and the precomputation advantage would carry over to those settings. Moreover, the reformulation of the tensor network contraction as a Grassmann integral over a quartic fermionic (pseudo) Hamiltonian \cite{wille2024minimaltensornetworkfree} opens up the possibility to employ methods from condensed matter theory, such as Feynman diagrams and self-consistent approximations. It is left for future work to evaluate the possible scope of the latter techniques for simulating quantum circuits. 

Further lines of research include the in-depth comparison to TN algorithms such as TEBD \cite{Vidal_2003}, fermionic TN contraction algorithms \cite{gao2024fermionictensornetworkcontraction} and the relation of the simulatability of circuits with our algorithm to their entanglement generation\cite{Projansky_2024_entanglement}, as well as the connection to learnability of fermionic states and Gaussian compressibility \cite{mele2024efficient}. It is also interesting to establish how our decomposition relates to the notion of fermionic nonlinearity studied in~\cite{Hakkaku_2022}.

Lastly, there have been recent systematic efforts to extend classical simulation techniques of Clifford and Matchgates to the so-called conjugated matchgate circuits~\cite{projansky2024}. It is interesting to see how our techniques could be extended to this setting and what computational overhead one would observe when simulating this class of circuits. 

\section{Acknowledgements}
C.W. was supported by the Royal Society Enhancement Award, S.S. was supported by the Royal Society University Research Fellowship and ``Quantum simulation algorithms for quantum chromodynamics'' grant (ST/W006251/1).

\bibliography{bibliography}

\begin{thebibliography}{40}%
\makeatletter
\providecommand \@ifxundefined [1]{%
 \@ifx{#1\undefined}
}%
\providecommand \@ifnum [1]{%
 \ifnum #1\expandafter \@firstoftwo
 \else \expandafter \@secondoftwo
 \fi
}%
\providecommand \@ifx [1]{%
 \ifx #1\expandafter \@firstoftwo
 \else \expandafter \@secondoftwo
 \fi
}%
\providecommand \natexlab [1]{#1}%
\providecommand \enquote  [1]{``#1''}%
\providecommand \bibnamefont  [1]{#1}%
\providecommand \bibfnamefont [1]{#1}%
\providecommand \citenamefont [1]{#1}%
\providecommand \href@noop [0]{\@secondoftwo}%
\providecommand \href [0]{\begingroup \@sanitize@url \@href}%
\providecommand \@href[1]{\@@startlink{#1}\@@href}%
\providecommand \@@href[1]{\endgroup#1\@@endlink}%
\providecommand \@sanitize@url [0]{\catcode `\\12\catcode `\$12\catcode `\&12\catcode `\#12\catcode `\^12\catcode `\_12\catcode `\%12\relax}%
\providecommand \@@startlink[1]{}%
\providecommand \@@endlink[0]{}%
\providecommand \url  [0]{\begingroup\@sanitize@url \@url }%
\providecommand \@url [1]{\endgroup\@href {#1}{\urlprefix }}%
\providecommand \urlprefix  [0]{URL }%
\providecommand \Eprint [0]{\href }%
\providecommand \doibase [0]{http://dx.doi.org/}%
\providecommand \selectlanguage [0]{\@gobble}%
\providecommand \bibinfo  [0]{\@secondoftwo}%
\providecommand \bibfield  [0]{\@secondoftwo}%
\providecommand \translation [1]{[#1]}%
\providecommand \BibitemOpen [0]{}%
\providecommand \bibitemStop [0]{}%
\providecommand \bibitemNoStop [0]{.\EOS\space}%
\providecommand \EOS [0]{\spacefactor3000\relax}%
\providecommand \BibitemShut  [1]{\csname bibitem#1\endcsname}%
\let\auto@bib@innerbib\@empty
\bibitem [{\citenamefont {Valiant}(2002)}]{Valiant}%
  \BibitemOpen
  \bibfield  {author} {\bibinfo {author} {\bibfnamefont {Leslie~G.}\ \bibnamefont {Valiant}},\ }\bibfield  {title} {\enquote {\bibinfo {title} {Quantum circuits that can be simulated classically in polynomial time},}\ }\href {\doibase 10.1137/S0097539700377025} {\bibfield  {journal} {\bibinfo  {journal} {SIAM Journal on Computing}\ }\textbf {\bibinfo {volume} {31}},\ \bibinfo {pages} {1229--1254} (\bibinfo {year} {2002})},\ \Eprint {http://arxiv.org/abs/https://doi.org/10.1137/S0097539700377025} {https://doi.org/10.1137/S0097539700377025} \BibitemShut {NoStop}%
\bibitem [{\citenamefont {Knill}(2001)}]{knill2001fermionic}%
  \BibitemOpen
  \bibfield  {author} {\bibinfo {author} {\bibfnamefont {E.}~\bibnamefont {Knill}},\ }\href {https://arxiv.org/abs/quant-ph/0108033} {\enquote {\bibinfo {title} {Fermionic linear optics and matchgates},}\ } (\bibinfo {year} {2001}),\ \Eprint {http://arxiv.org/abs/quant-ph/0108033} {arXiv:quant-ph/0108033 [quant-ph]} \BibitemShut {NoStop}%
\bibitem [{\citenamefont {Jozsa}\ and\ \citenamefont {Miyake}(2008)}]{jozsa2008matchgates}%
  \BibitemOpen
  \bibfield  {author} {\bibinfo {author} {\bibfnamefont {Richard}\ \bibnamefont {Jozsa}}\ and\ \bibinfo {author} {\bibfnamefont {Akimasa}\ \bibnamefont {Miyake}},\ }\bibfield  {title} {\enquote {\bibinfo {title} {Matchgates and classical simulation of quantum circuits},}\ }\href {\doibase 10.1098/rspa.2008.0189} {\bibfield  {journal} {\bibinfo  {journal} {Proceedings of the Royal Society A: Mathematical, Physical and Engineering Sciences}\ }\textbf {\bibinfo {volume} {464}},\ \bibinfo {pages} {3089–3106} (\bibinfo {year} {2008})}\BibitemShut {NoStop}%
\bibitem [{\citenamefont {Brod}\ and\ \citenamefont {Childs}(2013)}]{brod2013computational}%
  \BibitemOpen
  \bibfield  {author} {\bibinfo {author} {\bibfnamefont {Daniel~J}\ \bibnamefont {Brod}}\ and\ \bibinfo {author} {\bibfnamefont {Andrew~M}\ \bibnamefont {Childs}},\ }\bibfield  {title} {\enquote {\bibinfo {title} {The computational power of matchgates and the xy interaction on arbitrary graphs},}\ }\href {\doibase 10.26421/qic14.11-12} {\bibfield  {journal} {\bibinfo  {journal} {arXiv:1308.1463}\ }\textbf {\bibinfo {volume} {14}} (\bibinfo {year} {2013}),\ 10.26421/qic14.11-12}\BibitemShut {NoStop}%
\bibitem [{\citenamefont {Bravyi}\ and\ \citenamefont {Kitaev}(2002)}]{bravyi2002fermionic}%
  \BibitemOpen
  \bibfield  {author} {\bibinfo {author} {\bibfnamefont {Sergey~B.}\ \bibnamefont {Bravyi}}\ and\ \bibinfo {author} {\bibfnamefont {Alexei~Yu.}\ \bibnamefont {Kitaev}},\ }\bibfield  {title} {\enquote {\bibinfo {title} {Fermionic quantum computation},}\ }\href {\doibase 10.1006/aphy.2002.6254} {\bibfield  {journal} {\bibinfo  {journal} {Annals of Physics}\ }\textbf {\bibinfo {volume} {298}},\ \bibinfo {pages} {210–226} (\bibinfo {year} {2002})}\BibitemShut {NoStop}%
\bibitem [{\citenamefont {Bravyi}\ and\ \citenamefont {Koenig}(2011)}]{bravyi2011classical}%
  \BibitemOpen
  \bibfield  {author} {\bibinfo {author} {\bibfnamefont {Sergey}\ \bibnamefont {Bravyi}}\ and\ \bibinfo {author} {\bibfnamefont {Robert}\ \bibnamefont {Koenig}},\ }\href {https://arxiv.org/abs/1112.2184} {\enquote {\bibinfo {title} {Classical simulation of dissipative fermionic linear optics},}\ } (\bibinfo {year} {2011}),\ \Eprint {http://arxiv.org/abs/1112.2184} {arXiv:1112.2184 [quant-ph]} \BibitemShut {NoStop}%
\bibitem [{\citenamefont {Burkat}\ and\ \citenamefont {Strelchuk}(2024)}]{burkat2024lightweight}%
  \BibitemOpen
  \bibfield  {author} {\bibinfo {author} {\bibfnamefont {Jędrzej}\ \bibnamefont {Burkat}}\ and\ \bibinfo {author} {\bibfnamefont {Sergii}\ \bibnamefont {Strelchuk}},\ }\href {https://arxiv.org/abs/2404.07974} {\enquote {\bibinfo {title} {A lightweight protocol for matchgate fidelity estimation},}\ } (\bibinfo {year} {2024}),\ \Eprint {http://arxiv.org/abs/2404.07974} {arXiv:2404.07974 [quant-ph]} \BibitemShut {NoStop}%
\bibitem [{\citenamefont {Boixo}\ \emph {et~al.}(2018)\citenamefont {Boixo}, \citenamefont {Isakov}, \citenamefont {Smelyanskiy}, \citenamefont {Babbush}, \citenamefont {Ding}, \citenamefont {Jiang}, \citenamefont {Bremner}, \citenamefont {Martinis},\ and\ \citenamefont {Neven}}]{boixo2018characterizing}%
  \BibitemOpen
  \bibfield  {author} {\bibinfo {author} {\bibfnamefont {Sergio}\ \bibnamefont {Boixo}}, \bibinfo {author} {\bibfnamefont {Sergei~V.}\ \bibnamefont {Isakov}}, \bibinfo {author} {\bibfnamefont {Vadim~N.}\ \bibnamefont {Smelyanskiy}}, \bibinfo {author} {\bibfnamefont {Ryan}\ \bibnamefont {Babbush}}, \bibinfo {author} {\bibfnamefont {Nan}\ \bibnamefont {Ding}}, \bibinfo {author} {\bibfnamefont {Zhang}\ \bibnamefont {Jiang}}, \bibinfo {author} {\bibfnamefont {Michael~J.}\ \bibnamefont {Bremner}}, \bibinfo {author} {\bibfnamefont {John~M.}\ \bibnamefont {Martinis}}, \ and\ \bibinfo {author} {\bibfnamefont {Hartmut}\ \bibnamefont {Neven}},\ }\bibfield  {title} {\enquote {\bibinfo {title} {Characterizing quantum supremacy in near-term devices},}\ }\href {\doibase 10.1038/s41567-018-0124-x} {\bibfield  {journal} {\bibinfo  {journal} {Nature Physics}\ }\textbf {\bibinfo {volume} {14}},\ \bibinfo {pages} {595–600} (\bibinfo {year} {2018})}\BibitemShut {NoStop}%
\bibitem [{\citenamefont {Oszmaniec}\ \emph {et~al.}(2014)\citenamefont {Oszmaniec}, \citenamefont {Gutt},\ and\ \citenamefont {Kuś}}]{oszmaniec2014classical}%
  \BibitemOpen
  \bibfield  {author} {\bibinfo {author} {\bibfnamefont {Michał}\ \bibnamefont {Oszmaniec}}, \bibinfo {author} {\bibfnamefont {Jan}\ \bibnamefont {Gutt}}, \ and\ \bibinfo {author} {\bibfnamefont {Marek}\ \bibnamefont {Kuś}},\ }\bibfield  {title} {\enquote {\bibinfo {title} {Classical simulation of fermionic linear optics augmented with noisy ancillas},}\ }\href {\doibase 10.1103/physreva.90.020302} {\bibfield  {journal} {\bibinfo  {journal} {Physical Review A}\ }\textbf {\bibinfo {volume} {90}} (\bibinfo {year} {2014}),\ 10.1103/physreva.90.020302}\BibitemShut {NoStop}%
\bibitem [{\citenamefont {Brod}\ and\ \citenamefont {Galvão}(2011)}]{Brod_2011}%
  \BibitemOpen
  \bibfield  {author} {\bibinfo {author} {\bibfnamefont {Daniel~J.}\ \bibnamefont {Brod}}\ and\ \bibinfo {author} {\bibfnamefont {Ernesto~F.}\ \bibnamefont {Galvão}},\ }\bibfield  {title} {\enquote {\bibinfo {title} {Extending matchgates into universal quantum computation},}\ }\href {\doibase 10.1103/physreva.84.022310} {\bibfield  {journal} {\bibinfo  {journal} {Physical Review A}\ }\textbf {\bibinfo {volume} {84}} (\bibinfo {year} {2011}),\ 10.1103/physreva.84.022310}\BibitemShut {NoStop}%
\bibitem [{\citenamefont {Hebenstreit}\ \emph {et~al.}(2019)\citenamefont {Hebenstreit}, \citenamefont {Jozsa}, \citenamefont {Kraus}, \citenamefont {Strelchuk},\ and\ \citenamefont {Yoganathan}}]{all_magic}%
  \BibitemOpen
  \bibfield  {author} {\bibinfo {author} {\bibfnamefont {M.}~\bibnamefont {Hebenstreit}}, \bibinfo {author} {\bibfnamefont {R.}~\bibnamefont {Jozsa}}, \bibinfo {author} {\bibfnamefont {B.}~\bibnamefont {Kraus}}, \bibinfo {author} {\bibfnamefont {S.}~\bibnamefont {Strelchuk}}, \ and\ \bibinfo {author} {\bibfnamefont {M.}~\bibnamefont {Yoganathan}},\ }\bibfield  {title} {\enquote {\bibinfo {title} {All pure fermionic non-gaussian states are magic states for matchgate computations},}\ }\href {\doibase 10.1103/PhysRevLett.123.080503} {\bibfield  {journal} {\bibinfo  {journal} {Phys. Rev. Lett.}\ }\textbf {\bibinfo {volume} {123}},\ \bibinfo {pages} {080503} (\bibinfo {year} {2019})}\BibitemShut {NoStop}%
\bibitem [{\citenamefont {Cudby}\ and\ \citenamefont {Strelchuk}(2024)}]{cudby2023gaussian}%
  \BibitemOpen
  \bibfield  {author} {\bibinfo {author} {\bibfnamefont {Joshua}\ \bibnamefont {Cudby}}\ and\ \bibinfo {author} {\bibfnamefont {Sergii}\ \bibnamefont {Strelchuk}},\ }\href {https://arxiv.org/abs/2307.12654} {\enquote {\bibinfo {title} {Gaussian decomposition of magic states for matchgate computations},}\ } (\bibinfo {year} {2024}),\ \Eprint {http://arxiv.org/abs/2307.12654} {arXiv:2307.12654 [quant-ph]} \BibitemShut {NoStop}%
\bibitem [{Note1()}]{Note1}%
  \BibitemOpen
  \bibinfo {note} {The latter is defined for magic states \cite {all_magic} but can equally be applied to 2-qubit gates via Choi-Jamilkowski isomorphism.}\BibitemShut {Stop}%
\bibitem [{\citenamefont {Reardon-Smith}\ \emph {et~al.}(2024)\citenamefont {Reardon-Smith}, \citenamefont {Oszmaniec},\ and\ \citenamefont {Korzekwa}}]{ReardonSmith2024improvedsimulation}%
  \BibitemOpen
  \bibfield  {author} {\bibinfo {author} {\bibfnamefont {Oliver}\ \bibnamefont {Reardon-Smith}}, \bibinfo {author} {\bibfnamefont {Micha{\l{}}}\ \bibnamefont {Oszmaniec}}, \ and\ \bibinfo {author} {\bibfnamefont {Kamil}\ \bibnamefont {Korzekwa}},\ }\bibfield  {title} {\enquote {\bibinfo {title} {Improved simulation of quantum circuits dominated by free fermionic operations},}\ }\href {\doibase 10.22331/q-2024-12-04-1549} {\bibfield  {journal} {\bibinfo  {journal} {{Quantum}}\ }\textbf {\bibinfo {volume} {8}},\ \bibinfo {pages} {1549} (\bibinfo {year} {2024})}\BibitemShut {NoStop}%
\bibitem [{\citenamefont {Dias}\ and\ \citenamefont {Koenig}(2024)}]{dias2023classical}%
  \BibitemOpen
  \bibfield  {author} {\bibinfo {author} {\bibfnamefont {Beatriz}\ \bibnamefont {Dias}}\ and\ \bibinfo {author} {\bibfnamefont {Robert}\ \bibnamefont {Koenig}},\ }\bibfield  {title} {\enquote {\bibinfo {title} {Classical simulation of non-gaussian fermionic circuits},}\ }\href {\doibase 10.22331/q-2024-05-21-1350} {\bibfield  {journal} {\bibinfo  {journal} {Quantum}\ }\textbf {\bibinfo {volume} {8}},\ \bibinfo {pages} {1350} (\bibinfo {year} {2024})}\BibitemShut {NoStop}%
\bibitem [{\citenamefont {Gu}\ \emph {et~al.}(2010)\citenamefont {Gu}, \citenamefont {Verstraete},\ and\ \citenamefont {Wen}}]{gu2010grassmanntensornetworkstates}%
  \BibitemOpen
  \bibfield  {author} {\bibinfo {author} {\bibfnamefont {Zheng-Cheng}\ \bibnamefont {Gu}}, \bibinfo {author} {\bibfnamefont {Frank}\ \bibnamefont {Verstraete}}, \ and\ \bibinfo {author} {\bibfnamefont {Xiao-Gang}\ \bibnamefont {Wen}},\ }\href {https://arxiv.org/abs/1004.2563} {\enquote {\bibinfo {title} {Grassmann tensor network states and its renormalization for strongly correlated fermionic and bosonic states},}\ } (\bibinfo {year} {2010}),\ \Eprint {http://arxiv.org/abs/1004.2563} {arXiv:1004.2563 [cond-mat.str-el]} \BibitemShut {NoStop}%
\bibitem [{\citenamefont {Barthel}\ \emph {et~al.}(2009)\citenamefont {Barthel}, \citenamefont {Pineda},\ and\ \citenamefont {Eisert}}]{PhysRevA.80.042333}%
  \BibitemOpen
  \bibfield  {author} {\bibinfo {author} {\bibfnamefont {Thomas}\ \bibnamefont {Barthel}}, \bibinfo {author} {\bibfnamefont {Carlos}\ \bibnamefont {Pineda}}, \ and\ \bibinfo {author} {\bibfnamefont {Jens}\ \bibnamefont {Eisert}},\ }\bibfield  {title} {\enquote {\bibinfo {title} {Contraction of fermionic operator circuits and the simulation of strongly correlated fermions},}\ }\href {\doibase 10.1103/PhysRevA.80.042333} {\bibfield  {journal} {\bibinfo  {journal} {Phys. Rev. A}\ }\textbf {\bibinfo {volume} {80}},\ \bibinfo {pages} {042333} (\bibinfo {year} {2009})}\BibitemShut {NoStop}%
\bibitem [{\citenamefont {Kraus}\ \emph {et~al.}(2010)\citenamefont {Kraus}, \citenamefont {Schuch}, \citenamefont {Verstraete},\ and\ \citenamefont {Cirac}}]{PhysRevA.81.052338}%
  \BibitemOpen
  \bibfield  {author} {\bibinfo {author} {\bibfnamefont {Christina~V.}\ \bibnamefont {Kraus}}, \bibinfo {author} {\bibfnamefont {Norbert}\ \bibnamefont {Schuch}}, \bibinfo {author} {\bibfnamefont {Frank}\ \bibnamefont {Verstraete}}, \ and\ \bibinfo {author} {\bibfnamefont {J.~Ignacio}\ \bibnamefont {Cirac}},\ }\bibfield  {title} {\enquote {\bibinfo {title} {Fermionic projected entangled pair states},}\ }\href {\doibase 10.1103/PhysRevA.81.052338} {\bibfield  {journal} {\bibinfo  {journal} {Phys. Rev. A}\ }\textbf {\bibinfo {volume} {81}},\ \bibinfo {pages} {052338} (\bibinfo {year} {2010})}\BibitemShut {NoStop}%
\bibitem [{\citenamefont {Mortier}\ \emph {et~al.}(2025)\citenamefont {Mortier}, \citenamefont {Devos}, \citenamefont {Burgelman}, \citenamefont {Vanhecke}, \citenamefont {Bultinck}, \citenamefont {Verstraete}, \citenamefont {Haegeman},\ and\ \citenamefont {Vanderstraeten}}]{Mortier_2025}%
  \BibitemOpen
  \bibfield  {author} {\bibinfo {author} {\bibfnamefont {Quinten}\ \bibnamefont {Mortier}}, \bibinfo {author} {\bibfnamefont {Lukas}\ \bibnamefont {Devos}}, \bibinfo {author} {\bibfnamefont {Lander}\ \bibnamefont {Burgelman}}, \bibinfo {author} {\bibfnamefont {Bram}\ \bibnamefont {Vanhecke}}, \bibinfo {author} {\bibfnamefont {Nick}\ \bibnamefont {Bultinck}}, \bibinfo {author} {\bibfnamefont {Frank}\ \bibnamefont {Verstraete}}, \bibinfo {author} {\bibfnamefont {Jutho}\ \bibnamefont {Haegeman}}, \ and\ \bibinfo {author} {\bibfnamefont {Laurens}\ \bibnamefont {Vanderstraeten}},\ }\bibfield  {title} {\enquote {\bibinfo {title} {Fermionic tensor network methods},}\ }\href {\doibase 10.21468/scipostphys.18.1.012} {\bibfield  {journal} {\bibinfo  {journal} {SciPost Physics}\ }\textbf {\bibinfo {volume} {18}} (\bibinfo {year} {2025}),\ 10.21468/scipostphys.18.1.012}\BibitemShut {NoStop}%
\bibitem [{\citenamefont {Terhal}\ and\ \citenamefont {DiVincenzo}(2002)}]{Terhal_2002}%
  \BibitemOpen
  \bibfield  {author} {\bibinfo {author} {\bibfnamefont {Barbara~M.}\ \bibnamefont {Terhal}}\ and\ \bibinfo {author} {\bibfnamefont {David~P.}\ \bibnamefont {DiVincenzo}},\ }\bibfield  {title} {\enquote {\bibinfo {title} {Classical simulation of noninteracting-fermion quantum circuits},}\ }\href {\doibase 10.1103/physreva.65.032325} {\bibfield  {journal} {\bibinfo  {journal} {Physical Review A}\ }\textbf {\bibinfo {volume} {65}} (\bibinfo {year} {2002}),\ 10.1103/physreva.65.032325}\BibitemShut {NoStop}%
\bibitem [{\citenamefont {Bravyi}(2009)}]{Bravyi_2009}%
  \BibitemOpen
  \bibfield  {author} {\bibinfo {author} {\bibfnamefont {Sergey}\ \bibnamefont {Bravyi}},\ }\href {\doibase 10.1090/conm/482/09419} {\enquote {\bibinfo {title} {Contraction of matchgate tensor networks on non-planar graphs},}\ } (\bibinfo {year} {2009})\BibitemShut {NoStop}%
\bibitem [{\citenamefont {Brod}(2016)}]{Brod_2016}%
  \BibitemOpen
  \bibfield  {author} {\bibinfo {author} {\bibfnamefont {Daniel~J.}\ \bibnamefont {Brod}},\ }\bibfield  {title} {\enquote {\bibinfo {title} {Efficient classical simulation of matchgate circuits with generalized inputs and measurements},}\ }\href {\doibase 10.1103/physreva.93.062332} {\bibfield  {journal} {\bibinfo  {journal} {Physical Review A}\ }\textbf {\bibinfo {volume} {93}} (\bibinfo {year} {2016}),\ 10.1103/physreva.93.062332}\BibitemShut {NoStop}%
\bibitem [{\citenamefont {Lootens}\ \emph {et~al.}(2023)\citenamefont {Lootens}, \citenamefont {Delcamp}, \citenamefont {Ortiz},\ and\ \citenamefont {Verstraete}}]{Lootens_2023}%
  \BibitemOpen
  \bibfield  {author} {\bibinfo {author} {\bibfnamefont {Laurens}\ \bibnamefont {Lootens}}, \bibinfo {author} {\bibfnamefont {Clement}\ \bibnamefont {Delcamp}}, \bibinfo {author} {\bibfnamefont {Gerardo}\ \bibnamefont {Ortiz}}, \ and\ \bibinfo {author} {\bibfnamefont {Frank}\ \bibnamefont {Verstraete}},\ }\bibfield  {title} {\enquote {\bibinfo {title} {Dualities in one-dimensional quantum lattice models: Symmetric hamiltonians and matrix product operator intertwiners},}\ }\href {\doibase 10.1103/prxquantum.4.020357} {\bibfield  {journal} {\bibinfo  {journal} {PRX Quantum}\ }\textbf {\bibinfo {volume} {4}} (\bibinfo {year} {2023}),\ 10.1103/prxquantum.4.020357}\BibitemShut {NoStop}%
\bibitem [{\citenamefont {Wille}\ \emph {et~al.}(2024{\natexlab{a}})\citenamefont {Wille}, \citenamefont {Eisert},\ and\ \citenamefont {Altland}}]{Wille_2024}%
  \BibitemOpen
  \bibfield  {author} {\bibinfo {author} {\bibfnamefont {C.}~\bibnamefont {Wille}}, \bibinfo {author} {\bibfnamefont {J.}~\bibnamefont {Eisert}}, \ and\ \bibinfo {author} {\bibfnamefont {A.}~\bibnamefont {Altland}},\ }\bibfield  {title} {\enquote {\bibinfo {title} {Topological dualities via tensor networks},}\ }\href {\doibase 10.1103/physrevresearch.6.013302} {\bibfield  {journal} {\bibinfo  {journal} {Physical Review Research}\ }\textbf {\bibinfo {volume} {6}} (\bibinfo {year} {2024}{\natexlab{a}}),\ 10.1103/physrevresearch.6.013302}\BibitemShut {NoStop}%
\bibitem [{\citenamefont {O'Brien}\ \emph {et~al.}(2024)\citenamefont {O'Brien}, \citenamefont {Lootens},\ and\ \citenamefont {Verstraete}}]{Frank}%
  \BibitemOpen
  \bibfield  {author} {\bibinfo {author} {\bibfnamefont {Oliver}\ \bibnamefont {O'Brien}}, \bibinfo {author} {\bibfnamefont {Laurens}\ \bibnamefont {Lootens}}, \ and\ \bibinfo {author} {\bibfnamefont {Frank}\ \bibnamefont {Verstraete}},\ }\href {https://arxiv.org/abs/2404.07727} {\enquote {\bibinfo {title} {Local jordan-wigner transformations on the torus},}\ } (\bibinfo {year} {2024}),\ \Eprint {http://arxiv.org/abs/2404.07727} {arXiv:2404.07727 [quant-ph]} \BibitemShut {NoStop}%
\bibitem [{\citenamefont {Bravyi}(2004)}]{bravyi2004lagrangian}%
  \BibitemOpen
  \bibfield  {author} {\bibinfo {author} {\bibfnamefont {Sergey}\ \bibnamefont {Bravyi}},\ }\href {https://arxiv.org/abs/quant-ph/0404180} {\enquote {\bibinfo {title} {Lagrangian representation for fermionic linear optics},}\ } (\bibinfo {year} {2004}),\ \Eprint {http://arxiv.org/abs/quant-ph/0404180} {arXiv:quant-ph/0404180 [quant-ph]} \BibitemShut {NoStop}%
\bibitem [{Note2()}]{Note2}%
  \BibitemOpen
  \bibinfo {note} {If $G_{00,00}=0$ one needs to work with a slightly more complicated representation (cf. \cite {Bravyi_2009}), which is not discussed here.}\BibitemShut {Stop}%
\bibitem [{\citenamefont {Wille}\ \emph {et~al.}(2024{\natexlab{b}})\citenamefont {Wille}, \citenamefont {Usoltcev}, \citenamefont {Eisert},\ and\ \citenamefont {Altland}}]{wille2024minimaltensornetworkfree}%
  \BibitemOpen
  \bibfield  {author} {\bibinfo {author} {\bibfnamefont {Carolin}\ \bibnamefont {Wille}}, \bibinfo {author} {\bibfnamefont {Maksimilian}\ \bibnamefont {Usoltcev}}, \bibinfo {author} {\bibfnamefont {Jens}\ \bibnamefont {Eisert}}, \ and\ \bibinfo {author} {\bibfnamefont {Alexander}\ \bibnamefont {Altland}},\ }\href {https://arxiv.org/abs/2412.04216} {\enquote {\bibinfo {title} {A minimal tensor network beyond free fermions},}\ } (\bibinfo {year} {2024}{\natexlab{b}}),\ \Eprint {http://arxiv.org/abs/2412.04216} {arXiv:2412.04216 [cond-mat.str-el]} \BibitemShut {NoStop}%
\bibitem [{\citenamefont {Bravyi}\ and\ \citenamefont {Gosset}(2016)}]{PhysRevLett.116.250501}%
  \BibitemOpen
  \bibfield  {author} {\bibinfo {author} {\bibfnamefont {Sergey}\ \bibnamefont {Bravyi}}\ and\ \bibinfo {author} {\bibfnamefont {David}\ \bibnamefont {Gosset}},\ }\bibfield  {title} {\enquote {\bibinfo {title} {Improved classical simulation of quantum circuits dominated by clifford gates},}\ }\href {\doibase 10.1103/PhysRevLett.116.250501} {\bibfield  {journal} {\bibinfo  {journal} {Phys. Rev. Lett.}\ }\textbf {\bibinfo {volume} {116}},\ \bibinfo {pages} {250501} (\bibinfo {year} {2016})}\BibitemShut {NoStop}%
\bibitem [{Note3()}]{Note3}%
  \BibitemOpen
  \bibinfo {note} {The network for $p_M$ can be slightly simplified by removing all unitaries outside the $P_M$-light cone.}\BibitemShut {Stop}%
\bibitem [{\citenamefont {Wille}(2024)}]{PfExpansion}%
  \BibitemOpen
  \bibfield  {author} {\bibinfo {author} {\bibfnamefont {Carolin}\ \bibnamefont {Wille}},\ }\href {https://github.com/rho-tilde/BeyondMG} {\enquote {\bibinfo {title} {{Pfaffian Expansion}},}\ } (\bibinfo {year} {2024})\BibitemShut {NoStop}%
\bibitem [{\citenamefont {Wimmer}(2012)}]{Wimmer_2012}%
  \BibitemOpen
  \bibfield  {author} {\bibinfo {author} {\bibfnamefont {M.}~\bibnamefont {Wimmer}},\ }\bibfield  {title} {\enquote {\bibinfo {title} {Algorithm 923: Efficient numerical computation of the pfaffian for dense and banded skew-symmetric matrices},}\ }\href {\doibase 10.1145/2331130.2331138} {\bibfield  {journal} {\bibinfo  {journal} {ACM Transactions on Mathematical Software}\ }\textbf {\bibinfo {volume} {38}},\ \bibinfo {pages} {1–17} (\bibinfo {year} {2012})}\BibitemShut {NoStop}%
\bibitem [{\citenamefont {Hakkaku}\ \emph {et~al.}(2022)\citenamefont {Hakkaku}, \citenamefont {Tashima}, \citenamefont {Mitarai}, \citenamefont {Mizukami},\ and\ \citenamefont {Fujii}}]{Hakkaku_2022}%
  \BibitemOpen
  \bibfield  {author} {\bibinfo {author} {\bibfnamefont {Shigeo}\ \bibnamefont {Hakkaku}}, \bibinfo {author} {\bibfnamefont {Yuichiro}\ \bibnamefont {Tashima}}, \bibinfo {author} {\bibfnamefont {Kosuke}\ \bibnamefont {Mitarai}}, \bibinfo {author} {\bibfnamefont {Wataru}\ \bibnamefont {Mizukami}}, \ and\ \bibinfo {author} {\bibfnamefont {Keisuke}\ \bibnamefont {Fujii}},\ }\bibfield  {title} {\enquote {\bibinfo {title} {Quantifying fermionic nonlinearity of quantum circuits},}\ }\href {\doibase 10.1103/physrevresearch.4.043100} {\bibfield  {journal} {\bibinfo  {journal} {Physical Review Research}\ }\textbf {\bibinfo {volume} {4}} (\bibinfo {year} {2022}),\ 10.1103/physrevresearch.4.043100}\BibitemShut {NoStop}%
\bibitem [{\citenamefont {Mocherla}\ \emph {et~al.}(2023)\citenamefont {Mocherla}, \citenamefont {Lao},\ and\ \citenamefont {Browne}}]{mocherla2023extending}%
  \BibitemOpen
  \bibfield  {author} {\bibinfo {author} {\bibfnamefont {Avinash}\ \bibnamefont {Mocherla}}, \bibinfo {author} {\bibfnamefont {Lingling}\ \bibnamefont {Lao}}, \ and\ \bibinfo {author} {\bibfnamefont {Dan~E.}\ \bibnamefont {Browne}},\ }\href@noop {} {\enquote {\bibinfo {title} {Extending matchgate simulation methods to universal quantum circuits},}\ } (\bibinfo {year} {2023}),\ \Eprint {http://arxiv.org/abs/2302.02654} {arXiv:2302.02654 [quant-ph]} \BibitemShut {NoStop}%
\bibitem [{\citenamefont {Projansky}\ \emph {et~al.}(2024{\natexlab{a}})\citenamefont {Projansky}, \citenamefont {Necaise},\ and\ \citenamefont {Whitfield}}]{projansky2024}%
  \BibitemOpen
  \bibfield  {author} {\bibinfo {author} {\bibfnamefont {Andrew~M.}\ \bibnamefont {Projansky}}, \bibinfo {author} {\bibfnamefont {Jason}\ \bibnamefont {Necaise}}, \ and\ \bibinfo {author} {\bibfnamefont {James~D.}\ \bibnamefont {Whitfield}},\ }\href {https://arxiv.org/abs/2410.10068} {\enquote {\bibinfo {title} {Extending simulability of cliffords and matchgates},}\ } (\bibinfo {year} {2024}{\natexlab{a}}),\ \Eprint {http://arxiv.org/abs/2410.10068} {arXiv:2410.10068 [quant-ph]} \BibitemShut {NoStop}%
\bibitem [{\citenamefont {Miller}\ \emph {et~al.}(2025)\citenamefont {Miller}, \citenamefont {Holmes}, \citenamefont {Özlem Salehi}, \citenamefont {Chakraborty}, \citenamefont {Nykänen}, \citenamefont {Zimborás}, \citenamefont {Glos},\ and\ \citenamefont {García-Pérez}}]{miller2025simulationfermioniccircuitsusing}%
  \BibitemOpen
  \bibfield  {author} {\bibinfo {author} {\bibfnamefont {Aaron}\ \bibnamefont {Miller}}, \bibinfo {author} {\bibfnamefont {Zoë}\ \bibnamefont {Holmes}}, \bibinfo {author} {\bibnamefont {Özlem Salehi}}, \bibinfo {author} {\bibfnamefont {Rahul}\ \bibnamefont {Chakraborty}}, \bibinfo {author} {\bibfnamefont {Anton}\ \bibnamefont {Nykänen}}, \bibinfo {author} {\bibfnamefont {Zoltán}\ \bibnamefont {Zimborás}}, \bibinfo {author} {\bibfnamefont {Adam}\ \bibnamefont {Glos}}, \ and\ \bibinfo {author} {\bibfnamefont {Guillermo}\ \bibnamefont {García-Pérez}},\ }\href {https://arxiv.org/abs/2503.18939} {\enquote {\bibinfo {title} {Simulation of fermionic circuits using majorana propagation},}\ } (\bibinfo {year} {2025}),\ \Eprint {http://arxiv.org/abs/2503.18939} {arXiv:2503.18939 [quant-ph]} \BibitemShut {NoStop}%
\bibitem [{\citenamefont {Vidal}(2003)}]{Vidal_2003}%
  \BibitemOpen
  \bibfield  {author} {\bibinfo {author} {\bibfnamefont {Guifré}\ \bibnamefont {Vidal}},\ }\bibfield  {title} {\enquote {\bibinfo {title} {Efficient classical simulation of slightly entangled quantum computations},}\ }\href {\doibase 10.1103/physrevlett.91.147902} {\bibfield  {journal} {\bibinfo  {journal} {Physical Review Letters}\ }\textbf {\bibinfo {volume} {91}} (\bibinfo {year} {2003}),\ 10.1103/physrevlett.91.147902}\BibitemShut {NoStop}%
\bibitem [{\citenamefont {Gao}\ \emph {et~al.}(2024)\citenamefont {Gao}, \citenamefont {Zhai}, \citenamefont {Gray}, \citenamefont {Peng}, \citenamefont {Park}, \citenamefont {Liu}, \citenamefont {Kjønstad},\ and\ \citenamefont {Chan}}]{gao2024fermionictensornetworkcontraction}%
  \BibitemOpen
  \bibfield  {author} {\bibinfo {author} {\bibfnamefont {Yang}\ \bibnamefont {Gao}}, \bibinfo {author} {\bibfnamefont {Huanchen}\ \bibnamefont {Zhai}}, \bibinfo {author} {\bibfnamefont {Johnnie}\ \bibnamefont {Gray}}, \bibinfo {author} {\bibfnamefont {Ruojing}\ \bibnamefont {Peng}}, \bibinfo {author} {\bibfnamefont {Gunhee}\ \bibnamefont {Park}}, \bibinfo {author} {\bibfnamefont {Wen-Yuan}\ \bibnamefont {Liu}}, \bibinfo {author} {\bibfnamefont {Eirik~F.}\ \bibnamefont {Kjønstad}}, \ and\ \bibinfo {author} {\bibfnamefont {Garnet Kin-Lic}\ \bibnamefont {Chan}},\ }\href {https://arxiv.org/abs/2410.02215} {\enquote {\bibinfo {title} {Fermionic tensor network contraction for arbitrary geometries},}\ } (\bibinfo {year} {2024}),\ \Eprint {http://arxiv.org/abs/2410.02215} {arXiv:2410.02215 [quant-ph]} \BibitemShut {NoStop}%
\bibitem [{\citenamefont {Projansky}\ \emph {et~al.}(2024{\natexlab{b}})\citenamefont {Projansky}, \citenamefont {Heath},\ and\ \citenamefont {Whitfield}}]{Projansky_2024_entanglement}%
  \BibitemOpen
  \bibfield  {author} {\bibinfo {author} {\bibfnamefont {Andrew~M.}\ \bibnamefont {Projansky}}, \bibinfo {author} {\bibfnamefont {Joshuah~T.}\ \bibnamefont {Heath}}, \ and\ \bibinfo {author} {\bibfnamefont {James~D.}\ \bibnamefont {Whitfield}},\ }\bibfield  {title} {\enquote {\bibinfo {title} {Entanglement spectrum of matchgate circuits with universal and non-universal resources},}\ }\href {\doibase 10.22331/q-2024-08-07-1432} {\bibfield  {journal} {\bibinfo  {journal} {Quantum}\ }\textbf {\bibinfo {volume} {8}},\ \bibinfo {pages} {1432} (\bibinfo {year} {2024}{\natexlab{b}})}\BibitemShut {NoStop}%
\bibitem [{\citenamefont {Mele}\ and\ \citenamefont {Herasymenko}(2024)}]{mele2024efficient}%
  \BibitemOpen
  \bibfield  {author} {\bibinfo {author} {\bibfnamefont {Antonio~Anna}\ \bibnamefont {Mele}}\ and\ \bibinfo {author} {\bibfnamefont {Yaroslav}\ \bibnamefont {Herasymenko}},\ }\href@noop {} {\enquote {\bibinfo {title} {Efficient learning of quantum states prepared with few fermionic non-gaussian gates},}\ } (\bibinfo {year} {2024}),\ \Eprint {http://arxiv.org/abs/2402.18665} {arXiv:2402.18665 [quant-ph]} \BibitemShut {NoStop}%
\end{thebibliography}%

\appendix
\section{Mapping PPUs to fermionic tensors} \label{app:map}
Using the ordering of fermionic modes in Fig.~\ref{fig:map}, a PPU $G(a,b)$ parametrized as in Eq.~\eqref{eq:G} is mapped to a fermionic tensor $T=T_G + N \gamma \theta_1 \theta_2 \theta_3 \theta_4$ (cf. Eq.~\eqref{eq:Tdecomp}) with $N=a_{11}$, and $T_G=N e^{\frac 1 2 \theta^T A \theta}$ with
\begin{equation}
A= \frac{1}{N} \begin{pmatrix}
0 & a_{12} & b_{12} & b_{22}\\
-a_{12} & 0 & b_{11} & b_{21} \\
-b_{12} & -b_{11} & 0 & a_{21} \\
-b_{22}&-b_{21} &-a_{21} &0	
\end{pmatrix}\;, \\
 \end{equation}
and $\gamma={a_{22}}/{a_{11}}-\pf A $. 

\subsection{Explicit C-Phase decompositions}\label{app:As}
The Gaussian tensors for the C-Phase$(\phi)$ decompositions in Eq.~\eqref{eq:decomp} are given by
\begin{equation}
A_1(\phi)=   \begin{pmatrix}
0 & 0 & 0 & e^{\mathrm i \phi /2}\\
0 & 0 & e^{\mathrm i \phi /2} & 0 \\
0& -e^{\mathrm i \phi /2} & 0 &0\\
-e^{\mathrm i \phi /2}&0 &0 &0	
\end{pmatrix}\;,
 \end{equation}
 $A_2(\phi)=-A_1(\phi)$ and $A_\text{id}=A_1(\phi=0)$.

 \subsection{Consistent ordering}\label{app:JW}
\begin{figure}[h]
 	\centering
 	\includegraphics[width=0.25\textwidth]{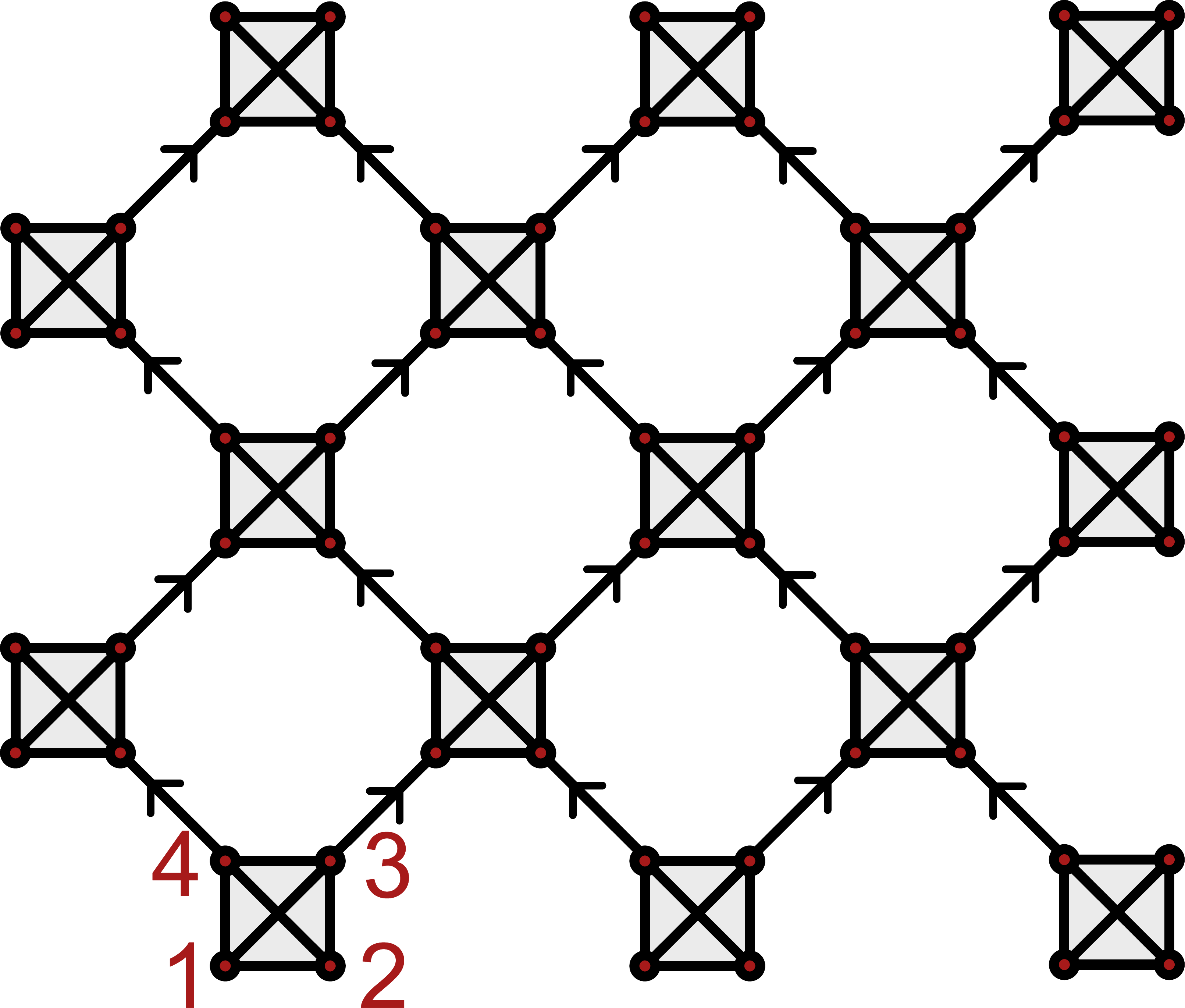}
 	\caption{Consistent assignment of fermionic mode ordering and contraction directions.}
 	\label{fig:lattice_arrows}
 \end{figure}

 When converting the bosonic tensor network to a fermionic tensor network, we want to ensure that the contraction commutes with fermionization. That is, for any tensor network we can either first fermionize the network and then contract it, or vice versa. This commutativity is a non-trivial condition. However, for any square lattice, it is not difficult to come up with a consistent assignment of orderings of the modes per tensor and directions of contractions such that commutativity is achieved. The concrete choice we use here is depicted in Fig.~\ref{fig:lattice_arrows}. For more details we refer to Appendix A of Ref.~\cite{Wille_2024}.
 
 \clearpage


\end{document}